\documentclass[aps,twocolumn,prb,amsmath,amssymb,superscriptaddress,longbibliography]{revtex4-2}
\usepackage{url}
\usepackage{graphicx}
\usepackage{color}
\usepackage{natbib}
\usepackage{hyperref}
\usepackage{notes2bib}
\usepackage{mathbbol}
\usepackage{float}
\usepackage{ulem}
\usepackage{tikz}
\usepackage{xcolor}
\usepackage{dsfont}
\usepackage{svg}

\newcommand{\bpm}{\begin{pmatrix}}
	\newcommand{\epm}{\end{pmatrix}}

\newcommand{\be}{\begin{equation}}
	\newcommand{\ee}{\end{equation}}
\newcommand{\beq}{\begin{eqnarray}}
	\newcommand{\eeq}{\end{eqnarray}}

\begin{document}

	\title{Quasiparticle interference and spectral function\\ of the UTe$_2$ superconductive surface band}
	\author{Adeline Cr\'epieux}
	\affiliation{Aix Marseille Univ, Universit\'e de Toulon, CNRS, CPT, Marseille, France}

        \author{Emile Pangburn}
        \affiliation{Institut de Physique Th\'eorique, Universit\'e Paris Saclay, CEA
		CNRS, Orme des Merisiers, 91190 Gif-sur-Yvette Cedex, France}
        \author{Shuqiu Wang}
        \affiliation{Clarendon Laboratory, University of Oxford, Oxford, 0X1 3PU, UK}
        \affiliation{H.H. Wills Physics Laboratory, University of Bristol, Bristol, BS8 1TL, UK}
        \author{Kuanysh Zhussupbekov}
        \affiliation{LASSP, Department of Physics, Cornell University, Ithaca, NY 14850, USA.}
        \affiliation{Department of Physics, University College Cork, Cork T12 R5C, IE}
        \author{Joseph~P.~Carroll}
        \affiliation{LASSP, Department of Physics, Cornell University, Ithaca, NY 14850, USA.}
        \affiliation{Department of Physics, University College Cork, Cork T12 R5C, IE}
        \author{Bin Hu}
        \affiliation{LASSP, Department of Physics, Cornell University, Ithaca, NY 14850, USA.}
        \author{Qiangqiang Gu}
        \affiliation{LASSP, Department of Physics, Cornell University, Ithaca, NY 14850, USA.}
        \author{J.C. S\'eamus Davis}
        \affiliation{LASSP, Department of Physics, Cornell University, Ithaca, NY 14850, USA.}
        \affiliation{Clarendon Laboratory, University of Oxford, Oxford, 0X1 3PU, UK}
        \affiliation{Department of Physics, University College Cork, Cork T12 R5C, IE}
        \affiliation{Max-Planck Institute for Chemical Physics of Solids, D-01187 Dresden, DE}
	\author{Catherine P\'epin}
	\affiliation{Institut de Physique Th\'eorique, Universit\'e Paris Saclay, CEA
		CNRS, Orme des Merisiers, 91190 Gif-sur-Yvette Cedex, France}
	\author{Cristina Bena}
	\affiliation{Institut de Physique Th\'eorique, Universit\'e Paris Saclay, CEA
		CNRS, Orme des Merisiers, 91190 Gif-sur-Yvette Cedex, France}
	
	\date{\today}

	\begin{abstract}
We compute the (0-11) surface spectral function, the surface density of states (DOS), and the quasiparticle interference (QPI) patterns, both in the normal state and superconducting (SC) state of UTe$_2$. We consider all possible non-chiral and chiral order parameters (OPs) that could in principle describe the superconductivity in this compound. We describe the formation of surface states whose maximum intensity energy depends on the nature of the pairing. We also study the QPI patterns resulting from the scattering of these surface states. Along the lines of \href{https://www.nature.com/articles/s41567-025-03000-w}{[Nat.~Phys.~21,~1555~(2025)]}, we show that the main feature distinguishing between various OPs is a QPI peak that is only observed experimentally in the superconducting state. The energy dispersion and the stability of this peak is consistent among the non-chiral OPs only with a $B_{3u}$ pairing. Moreover, $B_{3u}$ is the only non-chiral pairing that shows a peak at zero energy in the DOS, consistent with the experimental observations.
	\end{abstract}

\maketitle


\section{Introduction}

The heavy-fermion material uranium ditelluride (UTe$_2$) has recently been identified as a superconductor~\cite{aoki2019unconventional,ran2019nearly,knebel2019field,nakamine2019superconducting,tokunaga2019125te}, with a critical temperature of $T_c\sim 1.6-2.0$ K. Notably, unlike other uranium-based compounds~\cite{saxena2000superconductivity,aoki2001coexistence}, superconductivity in UTe$_2$ emerges from a paramagnetic normal state rather than from an ordered magnetic phase. Beyond its superconducting properties, UTe$_2$ displays other intriguing phenomena, including charge density wave (CDW) order~\cite{aishwarya2023magnetic,lafleur2024inhomogeneous}, which coexists with superconductivity—though this coexistence may be restricted to the (0-11) surface~\cite{theuss2024absence,kengle2024absence}. This interaction may lead to the formation of a modulated superconducting state~\cite{gu2023detection}, suggesting the possible emergence of a pair-density wave (PDW) state~\cite{aishwarya2023magnetic}.

Advancing our understanding of the underlying physics of UTe$_2$ relies on determining the symmetry of its superconducting order parameter (OP). While this is a well-defined and fundamental question, it has proven extremely challenging to resolve in non-BCS superconductors, as evidenced by the long-standing controversies surrounding Sr$_2$RuO$_4$~\cite{mackenzie2017even,pustogow2019constraints}. UTe$_2$ is widely believed to be a triplet superconductor, supported by various experimental observations, including an upper critical field $H_{c2}$ that significantly exceeds the Pauli limit~\cite{knebel2019field,ran2019nearly} and a minimal change in the Knight shift upon entering the superconducting phase~\cite{nakamine2019superconducting,ran2019nearly}, among other indications. However, more recent experiments suggest that these results are dependent on the crystallographic axis along which the magnetic field is applied~\cite{matsumura2023large}. The literature seems to converge on a $p$-wave-type pairing, with no argument for an $f$-wave pairing, consistent also with the fact that higher angular momentum pairings are in general energetically disfavored.

However, there is far less consensus regarding the specific odd-parity representation of the superconducting order parameter. 
Soon after the discovery of the superconducting phase, observations of two distinct superconducting transitions in the specific heat~\cite{hayes2021multicomponent},  power law dependence of the magnetic penetration depth~\cite{ishihara2023chiral}, along with a nonzero polar Kerr effect~\cite{hayes2021multicomponent} indicating time-reversal symmetry breaking (T), suggested that UTe$_2$ could be a two-component chiral superconductors~\cite{kittaka2020orientation} with chiral edge states~\cite{jiao2020chiral}. However, more recent measurements with better quality sample now seem to point to a single component superconductor. This is supported by recent reports showing no evidence of broken time-reversal symmetry, as indicated by the absence of a polar Kerr effect~\cite{ajeesh2023fate}, along with zero-energy Andreev peak measurements~\cite{gu2025pair}. Additional confirmation comes from ultrasound spectroscopy~\cite{theuss2024single} and SQUID measurements~\cite{iguchi2023microscopic}, reinforcing the case for a single-component superconducting order parameter. 

For a single-component order parameter, there are four possible pairing symmetries, corresponding to the odd-parity irreducible representations of the point group $D_{2h}$ of UTe$_2$, namely ${A_u, B_{1u}, B_{2u}, B_{3u}}$. The gapped or gapless nature of these superconducting order parameters depends on the topology of the normal-state Fermi surface of UTe$_2$, which remains a subject of active debate. Within the model considered in this work, the $A_u$ and $B_{1u}$ order parameters are fully gapped, whereas the $B_{2u}$ and $B_{3u}$ states exhibit gapless behavior. There is no consensus on which pairing symmetry is realized in UTe$_2$, with studies suggesting all possibilities, $A_u$~\cite{matsumura2023large, suetsugu2024fully}, $B_{1u}$\cite{iguchi2023microscopic}, $B_{2u}$\cite{theuss2024single}, and $B_{3u}$~\cite{fujibayashi2022superconducting, iguchi2023microscopic}.
The ongoing debate surrounding the determination of the superconducting order parameter symmetry in UTe$_2$, driven by various experimental observations, highlights the necessity of employing diverse experimental and theoretical approaches to make meaningful progress.

One popular approach that helps give insight into the nature of the superconducting order parameter is the study of the modifications to the density of states near an impurity. This can be directly probed using scanning tunneling microscopy (STM). The Fourier transforms of the real-space STM maps in the presence of impurities are also denoted quasiparticle interference (QPI) patterns. This method has been successfully applied to various superconductors, including $d$-wave cuprates~\cite{hudson2001interplay}, iron-pnictides~\cite{allan2012anisotropic,sprau2017discovery}, and strontium ruthenates~\cite{sharma2020momentum}. However, its main limitation, especially for bulk 3D materials, is that it serves only as a surface probe.

In this work, along the lines of Ref.~\onlinecite{experimentalpaper}, we investigate how distinct surface QPI patterns arising from different pairing symmetries can serve as a diagnostic tool to distinguish between various superconducting states in UTe$_2$. We use a recently developed Green's function technique~\cite{kaladzhyan2019obtaining,pinon2020surface} to compute the surface Green's function for the experimentally relevant (0-11) surface, as UTe$_2$ cleaves easily along this plane. Subsequently, using the well-established T-matrix formalism~\cite{balatsky2006impurity}, we compute the QPI patterns generated by a surface-localized impurity using the (0-11) surface Green's function. 

Several minimal models~\cite{ishizuka2021periodic,shishidou2021topological,tei2023possible} have been proposed based on ab-initio computations to explain the normal-state properties of UTe$_2$. It is widely believed that the Fermi surface of UTe$_2$ consists of two cylindrical sheets extending along the $c$-axis~\cite{xu2019quasi}, with only a weak dispersion in this direction as measured by magnetic quantum oscillations~\cite{eaton2024quasi}. Some authors have also proposed the presence of a small three-dimensional electron pocket originating from strongly correlated $f$-electrons, located either near the $\Gamma$-point~\cite{choi2022correlated,broyles2023revealing} or the $Z$-point~\cite{miao2020low}. This hypothesis is supported among others by core-level spectroscopy indirect evidence from transport measurements~\cite{Fujimori2019,fujimori2021core,eo2022c}. Although the presence of these Fermi pockets would influence the QPI interference pattern, since no definitive experimental signature has been observed to date, we will not explore this possibility in the current work.
Regarding the superconducting properties, although a single-component order parameter appears more likely, we focus on studying both single-component (${A_u, B_{1u}, B_{2u}, B_{3u}}$) and multi-component ($A_u+iB_{1u}$, $A_u+iB_{2u}$, $A_u+iB_{3u}$, $B_{1u}+iB_{2u}$, $B_{1u}+iB_{3u}$, $B_{2u}+iB_{3u}$) triplet pairing symmetries.

Although the main conclusions of our study have already been presented in Ref.~\onlinecite{experimentalpaper}, in the present work we provide a detailed development of the theoretical framework, as well as a comprehensive description of the theoretical analysis supporting the experimental STM results previously reported. Also, we extend the analysis of Ref.~\onlinecite{experimentalpaper}, which was focused mainly on the non-chiral $B_{2u}$ and $B_{3u}$ pairings to all the possible chiral and non-chiral pairing symmetries.

Our results indicate the formation of surface states for all the order parameters studied, however the only one showing a maximum of intensity at zero energy is $B_{3u}$, the rest appearing to be centered rather at finite energies, whose values depend on the form of the OP. Consequently, the surface DOS will have a peak at zero energy only for $B_{3u}$ and will show a double peak with a dip in the middle for all the rest of the OPs, however, due to the large quasiparticle damping, the split of the central peak will not be visible for all the OPs. All the surface states seem to have a maximum of intensity in the same regions in momentum space, following the normal state surface state pattern. However, at a given energy the resulting QPI differ greatly from one order parameter to the next. This is due first to the intensity distribution with energy which is different for each OP, but most importantly to the destructing interference stemming from phase cancellations which makes certain features disappear completely for the normal state or some of the OPs. We show that the spin structure of the surface states may play a role in these phase cancellations, however it is general hard to predict which wavevectors will be attenuated in the QPI patterns.

The most important conclusion of the comparison between our calculations and the experimental measurements is that among the non-chiral OPs, which are believed to be the most likely to describe the UTe$_2$ physics, the only one consistent with the measurements is $B_{3u}$. The first argument to this effect is the surface state DOS which is showing a peak in the experimental observations. Secondly, while many of the QPI peaks observed experimentally are common to different OPs, and some even to the normal state, there is one peak whose presence and stability cannot be explained except by the $B_{3u}$ order parameter.

The structure of the paper is the following. In Section \ref{Sec:Model}, we outline the tight-binding model and the method used to compute the surface spectral function and the QPI patterns for an impurity located on the (0-11) surface of UTe$_2$. In Sections \ref{Sec:ResultsNormal} and \ref{Sec:Results}, we present the results obtained for UTe$_2$ in its normal state, and respectively in its superconducting state, for both non-chiral and chiral pairings. We discuss these results and conclude in Section~\ref{Sec:Discussions}. Additional details on the model and method are provided in Appendix~\ref{App:ModelMethod}, and a detailed discussion on symmetry constraints is given in Appendix~\ref{App:Symmetry_Edge}. In Appendices~ \ref{App:PlotEzero}  and \ref{App:E_Plot} we provide the plots for the bulk spectral function and the JDOS at zero-energy, as well as the plots for the surface spectral function and the QPI patterns at finite energies.


\section{Model and methods\label{Sec:Model}}

\subsection{Tight-binding model\label{Sec:TB_Model}}

\begin{figure}[h!]
\includegraphics[width=4cm]{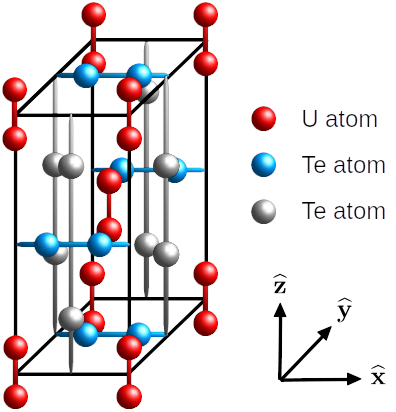}
\caption{Schematic picture of the UTe$_2$ lattice. The 4-orbital tight-binding model used in this work includes nearest-neighbor hopping 
between U~atoms, nearest-neighbor hopping 
between blue color Te~atoms, as gray color  Te~atoms weakly affect the Fermi surface\cite{theuss2024single}, and hybridization $\delta$ between U and Te energy bands.}
\label{Fig:Lattice}
\end{figure}

\begin{figure}[h!]
\includegraphics[width=4.3cm]{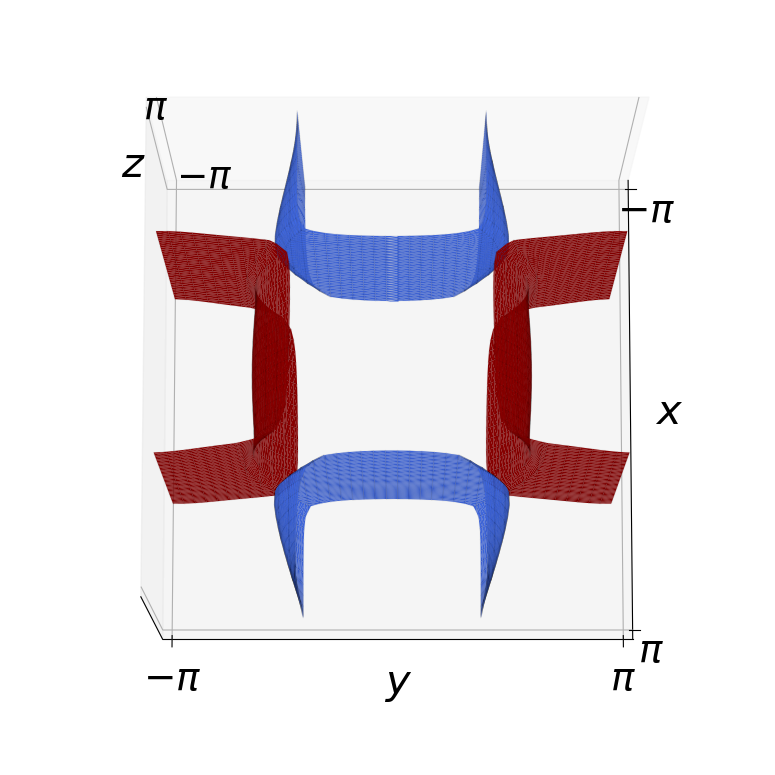}
\includegraphics[width=4.2cm]{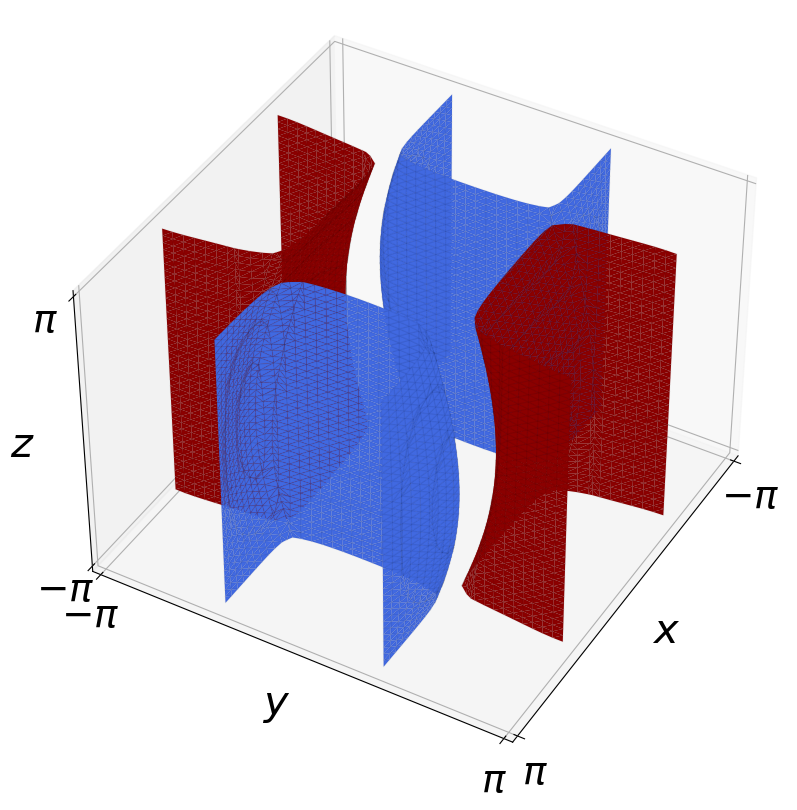}
\caption{Top view and 3D view of Fermi surface in the 4-orbital model in the presence of hybridization $\delta$.}\label{Fig:NS_FermiSurface}
\end{figure}

UTe$_2$ has a body-centered orthorhombic lattice structure with the space group symmetry Immm~\cite{bradley2009mathematical} and lattice constants $a = 0.41$ nm, $b = 0.61$ nm and $c = 1.39$ nm for the dimensions of the unit cell along $\widehat{x}$-, $\widehat{y}$- and $\widehat{z}$-axis (see Fig.~\ref{Fig:Lattice}). We use a four-orbital tight-binding model based on density functional theory (DFT) calculations introduced by Theuss et al.~\cite{theuss2024single}. Two sets of parameters are given by these authors: one matching DFT results and the other matching quantum-oscillation (QO) experiments. Our approach was to start with the DFT tight-binding parameters and modify their values (see Appendix \ref{App:TBparameters}) to match the experimental constraints on the Fermi surface. This model, which exhibits two cylindrical-like Fermi surface sheets, does not consider the hypothesized Fermi pocket centered around $\Gamma$ predicted by some authors based on strongly correlated physics beyond DFT-based approaches~\cite{choi2022correlated}, due to conflicting experimental evidence~\cite{broyles2023revealing,eaton2024quasi}. The presence of two uranium atoms and two tellurium atoms per unit cell leads to four bands of energy described by the Hamiltonian written in orbital space $\mathcal{E}_{orb}$
\begin{eqnarray}
\mathcal{H}_\mathrm{TB}(\mathbf{k})=\begin{pmatrix}
\mathcal{H}_\mathrm{U}(\mathbf{k}) & \mathcal{H}_\delta \\
\mathcal{H}_\delta^\dagger & \mathcal{H}_\mathrm{Te}(\mathbf{k}) \\
\end{pmatrix}&~,
\label{Eq:TB_model}
\end{eqnarray}
where
\begin{eqnarray}
\mathcal{H}_\mathrm{U}(\mathbf{k})&=&
\begin{pmatrix}
\mathcal{E}_\mathrm{U}(\mathbf{k}) & f_\mathrm{U}(\mathbf{k}) \\
f_\mathrm{U}^*(\mathbf{k}) & \mathcal{E}_\mathrm{U}(\mathbf{k}) \\
\end{pmatrix}~,\\
\mathcal{H}_\mathrm{Te}(\mathbf{k})&=&
\begin{pmatrix}
\mathcal{E}_\mathrm{Te}(\mathbf{k}) & f_\mathrm{Te}(\mathbf{k}) \\
f_\mathrm{Te}^*(\mathbf{k}) & \mathcal{E}_\mathrm{Te}(\mathbf{k}) \\
\end{pmatrix}~,
\end{eqnarray}
with
\begin{eqnarray}
\mathcal{E}_\mathrm{U}(\mathbf{k})&=&\mu_U-2t_U\cos(ak_x)-2t_{ch,U}\cos(bk_y)~, \\
f_\mathrm{U}(\mathbf{k})&=&-\Delta_U-2t'_U\cos(ak_x)-2t'_{ch,U}\cos(bk_y)\nonumber\\
&-&4t_{z,U}e^{-ick_z/2}\cos(ak_x/2)\cos(bk_y/2)~,\\
\mathcal{E}_\mathrm{Te}(\mathbf{k})&=&\mu_{T_e}-2t_{ch,T_e}\cos(ak_x)~,\\
f_\mathrm{Te}(\mathbf{k})&=&-\Delta_{T_e}-t_{T_e}\exp(-ibk_y)\nonumber\\
&-&2t_{z,T_e}\cos(ck_z/2)\cos(ak_x/2)\cos(bk_y/2)~,
\end{eqnarray}
and
\begin{eqnarray}
\mathcal{H}_\delta=
\begin{pmatrix}
\delta & 0 \\
0 & \delta \\
\end{pmatrix}~,
\end{eqnarray}
with $\delta$ the hybridization strength between $U$ and $Te$ atoms. In the absence of hybridization, i.e. for $\delta=0$, two of the four bands cross the Fermi energy. Introducing a hybridization creates electron and hole pockets with cylindrical-like shapes aligned along the $\widehat{z}$-axis. We verified that the specific form of the hybridization is not crucial as long as it results in two disconnected pockets. We chose the value of the hybridization parameter~$\delta$ such that the area of the electron and hole pockets at $k_z=0$ matches recent quantum-oscillation experiments~\cite{eaton2024quasi,weinberger2024quantum}, with $\mathcal{A}_e\approx\mathcal{A}_h\approx \ 33.6$ nm$^{2}$. The resulting Fermi surface is shown in Fig.~\ref{Fig:NS_FermiSurface}. The main characteristic is the presence of weakly dispersing cylindrical-like Fermi pockets parallel to the $\widehat{z}$-axis . We see moreover that this Fermi surface is symmetrical under the Immm space group symmetries.

\subsection{Expression for the gap function}

In this work we restrict the analysis to a triplet $p$-wave superconducting whose possible physical origin is the pairing driven by a ferromagnetic quantum critical point~\cite{ran2019nearly,jiao2020chiral}. Pairing symmetries in a crystal are classified by the point group of the model: for UTe$_2$ the point group is $D_{2h}$~\cite{tei2023possible}. Because inversion symmetry $\mathcal{I}\in D_{2h}$, the triplet pairing can be classified either as odd or as even under inversion. The gap function for a triplet superconductor in spin space can be written as:
\begin{eqnarray}
&\Delta(\mathbf{k})=\left[{\bf d}(\mathbf{k})\cdot \boldsymbol{\sigma}\right]\left(i\sigma_y\right)~,&
\label{Eq:SC_OP}
\end{eqnarray}
with $\boldsymbol{\sigma}=\left\{\sigma_x,\sigma_y,\sigma_z\right\}$ the Pauli matrix, and ${\bf d}(\mathbf{k})$ a vector which characterizes the pairing. In single-orbital superconductors, the Pauli principle implies that the gap function should satisfy the relation $\Delta(\mathbf{k})=-\Delta^T(-\mathbf{k})$. For more complicated orbital pairings, such as those we study here, $\Delta(\mathbf{k})$ can also be non-trivial in orbital space. Because the electrons contributing to the Fermi surface are $f$-electrons, they experience a strong spin-orbit coupling. Consequently, point group transformations and spin transformations are no longer independent: acting with certain symmetries on the lattice results in corresponding actions on the spin~\cite{sigrist1991phenomenological}. This implies that the direction of the ${\bf d}(\mathbf{k})$ vector is constrained by the point group representation $D_{2h}$ which has eight irreducible representations: $A_{g/u}$, $B_{1g/u}$, $B_{2g/u}$ and $B_{3g/u}$~\cite{aroyo2006bilbao}. However, since we consider triplet pairing, it should be odd under inversion~$\mathcal{I}$, such that the allowed representations are $A_u$, $B_{1u}$, $B_{2u}$ and $B_{3u}$. All these representations correspond to non-chiral order parameters invariant under the time-reversal symmetry represented by $\mathcal{T}=i\sigma_y\mathcal{K}$, where $\mathcal{K}$ is complex conjugation. Chiral order parameters~\cite{kallin2016chiral} can be constructed by adding two representations which are energetically degenerate with a relative phase of $\pi/ 2$ such that it breaks the time-reversal symmetry~\cite{sato2017topological}. Since we are not solving the self-consistent BCS gap equations to compare the free energies, we consider different combinations, even if they may be disfavored in reality. The transformation properties of these pairing symmetries are listed in Tab.~\ref{Table:RepSC}.

Because there are two uranium orbitals and two tellurium orbitals in our tight-binding model, we can have both intra- and inter-orbital pairing. In the following, we only consider inter-orbital pairing states with orbital-triplet pairings~\cite{tei2023possible} and restrict the pairings to nearest-neighbor couplings along the $\widehat{x}$-, $\widehat{y}$- or $\widehat{z}$-axis. We, moreover, assume that the pairing strength is independent of the axis direction. This leads to the following form for the non-chiral ${d}$-vector
\begin{eqnarray}
{\bf d}_{A_{u}}(\mathbf{k})&=&\Delta_0\,\Big(\sin(ak_x),\sin(bk_y),\sin(ck_z)\Big)^T~,\\
{\bf d}_{B_{1u}}(\mathbf{k})&=&\Delta_0\,\Big(\sin(bk_y),\sin(ak_x),0\Big)^T~,\\
{\bf d}_{B_{2u}}(\mathbf{k})&=&\Delta_0\,\Big(\sin(ck_z),0,\sin(ak_x)\Big)^T~,\\
\label{eq:db3u}{\bf d}_{B_{3u}}(\mathbf{k})&=&\Delta_0\,\Big(0,\sin(ck_z),\sin(bk_y)\Big)^T~,
\end{eqnarray}
where $\Delta_0$ is the pairing strength. The form of the chiral ${d}$-vectors follows readily by adding two non-chiral ${d}$-vectors with a relative phase of $\pi/2$. Note that the point group $D_{2h}$ does not impose that pairing strength $\Delta_0$ has to be independent of the axis direction, meaning that one can relax this assumption and take a more general $\mathbf {d}(\mathbf{k})$ if needed. Others terms are also allowed by symmetry for the ${d}$-vectors~\cite{hillier2012nonunitary}. For instance, the $d$-vector for $B_{3u}$ pairing can be generalized to $
\mathbf{d}_{B_{3u}}(\mathbf{ k})=\left(C_0\sin(ak_x)\sin(bk_y)\sin(ck_z),\Delta_0\sin(ck_z),\Delta_0\sin(bk_y)\right)^T$. The QPI results for $C_0\leq \Delta_0$ are qualitatively unchanged.

\begin{table}[h!]
\centering
\begin{tabular}{|c|c|c|c|c|c|c|}
\hline
IR & $C_z$ & $C_y$ & $C_x$ & $M_z$ & $M_y$ & $M_x$ \\ 
\hline
\hline
$A_u$  & 1 & 1 & 1 & -1 & -1 & -1 \\
\hline 
$B_{1u}$  & 1 & -1 & -1 & -1 & 1 & 1 \\
\hline
$B_{2u}$  & -1 & 1 & -1 & 1 & -1 & 1 \\
\hline
$B_{3u}$  & -1 & -1 & 1 & 1 & 1 & -1 \\
\hline
$A_u+iB_{1u}$  & 1 & $\times$ & $\times$ & -1 & $\times$ & $\times$ \\
\hline
$A_u+iB_{2u}$  & $\times$ & 1 & $\times$ & $\times$ & -1 & $\times$ \\
\hline
$A_u+iB_{3u}$  & $\times$ & $\times$ & 1 & $\times$ & $\times$ & -1 \\
\hline
$B_{1u}+iB_{2u}$  & $\times$ & $\times$ & -1 & $\times$ & $\times$ & 1 \\
\hline
$B_{1u}+iB_{3u}$  & $\times$ & -1 & $\times$ & $\times$ & 1 & $\times$ \\
\hline
$B_{2u}+iB_{3u}$  & -1 & $\times$ & $\times$ & 1 & $\times$ & $\times$ \\
\hline
\end{tabular}
\caption{List of irreducible representations of $D_{2h}$ and symmetry transformations for each pairing: $C_{\alpha}$ and  $M_{\alpha}$ respectively refer to a rotation with angle~$\pi$ around $\widetilde \alpha$-axis and a mirror reflection by a plane normal to the $\widetilde \alpha$-axis, where $\alpha=x,y,z$. The symbol~$\times$ indicates that the ${\bf d}(\mathbf{k})$ vector breaks the symmetry.}
\label{Table:RepSC}
\end{table}

\subsection{Surface Green function calculation\label{sec:Surface_Green}}

The $T$-matrix method is an exact analytical method to compute Green's function for a $\delta$-localized impurity in a non-interacting infinite system~\cite{balatsky2006impurity}.  This method can be extended to compute the boundary states of a $d$-dimensional bulk Hamiltonian by considering a ($d$-1)-dimensional localized impurity, which effectively splits the system in half~\cite{kaladzhyan2019obtaining,pinon2020surface} for a very large impurity potential. For example, for a 3D system, the Green's function computed near a plane-like impurity will converge to the surface Green function in the limit of a large impurity potential. 

The Green function of the clean Bogoliubov-de-Gennes Hamiltonian $\mathcal{H}_\mathrm{BdG}(\mathbf{k})$, built from $\mathcal{H}_\mathrm{TB}(\mathbf{k})$ and $\Delta(\mathbf{k})$ and given by Eq.~(\ref{Eq:BGH_model}), can be written as:
\begin{eqnarray}\label{Eq:BulkGreenFunction}
&G_\mathrm{BdG}(E,\mathbf{k})=\Big((E+i\eta)\mathbb{1}_{16}-\mathcal{H}_\mathrm{BdG}(\mathbf{k})\Big)^{-1}~,&
\end{eqnarray}
with  $\eta$ the quasiparticle damping. $\mathcal{H}_\mathrm{BdG}(\mathbf{k})$ is a $16\times 16$ matrix in the Hilbert space $\mathcal{E}=\mathcal{E}_{orb} \otimes \mathcal{E}_{spin}\otimes \mathcal{E}_{el-h}$ with $\mathcal{E}_{orb}$ the orbital subspace $\mathcal{E}_{spin}$ the spin subspace and $\mathcal{E}_{el-h}$ the electron-hole subspace (see Appendix~\ref{App:HBdG}). $\mathbb{1}_n$ is an $n \times n$ identity matrix. To compute surface states in the cleave plane (0-11), we consider an impurity plane such that translation invariance is maintained in direction~$\mathbf{k}_\parallel$ parallel to the plane and broken in orthogonal direction~$\mathbf{k}_\perp$. By decomposing~$\mathbf{r}$ into vectors perpendicular and parallel to the impurity plane, $\mathbf{r}=(\mathbf{r}_\parallel,\mathbf{r}_\perp)$, the impurity potential is given by $V(\mathbf{r})=\mathds{V}\delta(\mathbf{r}_\perp)$, with
\begin{eqnarray}
\mathds{V}=V_0
\begin{pmatrix}
\mathds{1}_8 & 0 \\
0 & -\mathds{1}_8 \\
\end{pmatrix}~,
\end{eqnarray}
where $V_0$ is the plane impurity strength. We make the choice to locate the impurity plane at position $\mathbf{r}_\perp=0$. For such an extended impurity, the T-matrix expression is
\begin{eqnarray}
\label{Eq:Tmatrix_Surface}
&&T(E,\mathbf{k}_{\parallel})=\left(\mathds{1}_{16}-\mathds{V}\int_{BZ_\perp} \dfrac{\mathrm{d}\mathbf{k}_\perp}{L_{BZ_\perp}}G_\mathrm{BdG}(E,\mathbf{k}_\perp,\mathbf{k}_{\parallel})\right)^{-1}\mathds{V}~,\nonumber\\
\end{eqnarray}
where $L_{BZ_\perp}$ is the length of the Brillouin zone along the axis perpendicular to the (0-11) plane, and where we have introduced $\mathbf{k}=(\mathbf{k}_\parallel,\mathbf{k}_\perp)$. The technical details on the $\mathbf{k}_\perp$ integration are given in Appendix~\ref{App:MethodSGF}. From the knowledge of the T-matrix, the Green function in the presence of the impurity plane can be exactly computed since one has
\begin{eqnarray}
\label{Eq:SurfaceGreenFunction}
&&G(E,\mathbf{k}_{\perp 1},\mathbf{k}_{\perp 2},\mathbf{k}_{\parallel})=G_\mathrm{BdG}(E,\mathbf{k}_{\perp 1},\mathbf{k}_\parallel)\delta({\mathbf{k}_{\perp1}-\mathbf{k}_{\perp2}})\nonumber\\
&&+G_\mathrm{BdG}(E,\mathbf{k}_{\perp 1},\mathbf{k}_{\parallel})T(E,\mathbf{k}_{\parallel})G_\mathrm{BdG}(E,\mathbf{k}_{\perp2},\mathbf{k}_{\parallel})~.
\end{eqnarray}
The surface Green function $G_s(E,\mathbf{k}_\parallel)$ is defined as the double Fourier transform of the Green function one-lattice spacing away from the impurity plane in the perpendicular direction, that is
\begin{eqnarray}
G_s(E,\mathbf{k}_\parallel)&=&\int_{BZ_\perp} \int_{BZ_\perp}\dfrac{\mathrm{d}\mathbf{k}_{\perp 1}\mathrm{d}\mathbf{k}_{\perp 2}}{L^2_{BZ_\perp}}e^{id_\perp\left(\mathbf{k}_{\perp 1}-\mathbf{k}_{\perp 2}\right)\cdot \mathbf{e}_\perp}\nonumber\\
&&\times G(E,\mathbf{k}_ {\perp 1},\mathbf{k}_{\perp 2},\mathbf{k}_\parallel)~,
\end{eqnarray}
with $d_\perp=bc/\sqrt{b^2+c^2}$ is the distance between the (0-11) impurity plane and its nearest lattice plane. Using Eq.~\ref{Eq:SurfaceGreenFunction}, we can decompose $G_s(E, \mathbf{k}_\parallel)$ into a sum of two terms
\begin{eqnarray}\label{Eq:GS_Equation}
G_s(E,\mathbf{k}_\parallel)=G_{b}(E,\mathbf{k}_\parallel,0)+G_{i}(E,\mathbf{k}_\parallel)~,
\end{eqnarray}
where $G_b$ and $G_{i}$ correspond to the ``bulk'' contribution and to the impurity contribution, and are given by
\begin{eqnarray}
\label{Eq:BulkGreen}
G_b(E,\mathbf{k}_\parallel,z_\perp)&=&\int_{BZ_\perp} \dfrac{\mathrm{d}\mathbf{k}_\perp}{L_{BZ_\perp}}e^{i z_\perp\mathbf{k}_\perp\cdot\mathbf{e}_\perp}G_\mathrm{BdG}(E,\mathbf{k}_\perp,\mathbf{k}_\parallel)~,\nonumber\\~\\
\label{Eq:EdgeGreen}
G_{i}(E,\mathbf{k}_\parallel)&=&G_b(E,\mathbf{k}_\parallel,d_\perp)T(E,\mathbf{k}_\parallel)G_b(E,\mathbf{k}_\parallel,-d_\perp)~.\nonumber\\
\end{eqnarray}
We have chosen an integration region with the same area as the first BZ, but of a rectangular shape, following Ref.~\cite{pinon2020surface}. Due to the $k$-space periodicity, this is equivalent to integrating over the first BZ and much more straightforward. We note that the bulk contribution corresponds solely to a simple projection of the 3D-bulk physics on the (0-11) surface, without taking into account the existence of a surface and the semi-infinite nature of the system. The surface physics is hidden in the impurity term $G_{i}(E,\mathbf{k}_\parallel)$ which when the impurity potential is taken to infinity will describe the formation of the surface states.

The surface Green's function can subsequently be used to calculate the surface spectral function:
\begin{eqnarray}
A_s(E,\mathbf{k}_\parallel)=-\frac{1}{\pi}\mathrm{Im}\{\mathrm{Tr}_\mathrm{el}[G_s(E,\mathbf{k}_\parallel)]\}~,
 \label{Eq:SurfaceSpectral}
\end{eqnarray}
where the trace runs over the electron bands. We can also here distinguish two components, and in what follows we sometimes calculate separately the  ``bulk'' spectral function
\begin{eqnarray}
A_b(E,\mathbf{k}_\parallel)=-\frac{1}{\pi}\mathrm{Im}\{\mathrm{Tr}_\mathrm{el}[G_b(E,\mathbf{k}_\parallel,0)]\}~,
 \label{Eq:BulkSpectral}
\end{eqnarray}
which is just the component of the surface spectral function arising from the projection of the 3D-bulk spectral function on the (0-11) surface. Once more, this component will not show any of the novel physics associated with the semi-infinite character of the system.

\subsection{QPI calculation\label{Sec:QPI_Method}}

The QPI pattern corresponds to the Fourier transform of the local density of states at a specific energy. We compute the QPI pattern resulting from the physical scenario of an impurity located on the (0-11) surface. This can then be directly compared with STM experiments~\cite{gu2023detection,gu2025pair}, taking into account all the surface effects. To compute the QPI for such a point-like impurity, the surface Green function $G_s(E,\mathbf{k}_\parallel)$ has first to be computed using Eq.~(\ref{Eq:GS_Equation}).  Second, one has to compute the T-matrix associated with the point-like impurity~\cite{balatsky2006impurity}, whose poles correspond to impurity states
\begin{eqnarray}
\label{Eq:Tmatrix}T_{s}(E)=\left(\mathds{1}-\mathds{U}_{s}\int_{BZ_\parallel} \dfrac{\mathrm{d}^2\mathbf{k}_{\parallel}}{S_{BZ_\parallel}} G_{s}(E,\mathbf{k}_{\parallel})\right)^{-1}\mathds{U}_{s}~,
\end{eqnarray}
where $S_{BZ}$ is the first Brillouin zone area in the (0-11) plane, and where the impurity matrix is
\begin{eqnarray}
\mathds{U}_s=U_0
\begin{pmatrix}
\mathds{1}_8 & 0 \\
0 & -\mathds{1}_8 \\
\end{pmatrix}~,&
\end{eqnarray}
with $U_0$ the strength of the point-like impurity. The scatterer considered in our analysis is non-magnetic and
has equal components on all four orbitals. We checked that QPI features are qualitatively unchanged for different forms of the potential. The physical observables, such as the local density of states (LDOS) that can be measured near an impurity, can be expressed directly in terms of this T-matrix, if we assume the dilute-limit approximation, such that the impurities are well separated from each other. The Fourier transform of the change in LDOS induced by the impurity, $\delta \rho (E,\mathbf{q}_\parallel)$, can then be written as
\begin{eqnarray}\label{Eq:QPI}
\delta \rho(E,\mathbf{q}_\parallel)=-\dfrac{1}{2\pi i}\int_{BZ_\parallel} \dfrac{\mathrm{d}^2 \mathbf{k}_\parallel}{S_{BZ_\parallel}}
\mathrm{Tr}_\mathrm{el}\left[\,\widetilde{g}(E,\mathbf{k}_\parallel,\mathbf{q}_\parallel)\right]~,\nonumber  \\
\end{eqnarray}
where
\begin{eqnarray}
\label{Eq:LDOS}
\widetilde{g}(E,\mathbf{k}_\parallel,\mathbf{q}_\parallel)&=&{G}_s(E,\mathbf{k}_\parallel)T_s(E){G}_s(E,\mathbf{k}_\parallel+\mathbf{q}_\parallel)\nonumber  \\
 & -& {G}_s^{\ast}(E,\mathbf{k}_\parallel+\mathbf{q}_\parallel)T_s^{\ast}(E){G}_s^{\ast}(E,\mathbf{k}_\parallel)~.
\end{eqnarray}
At $\mathbf{q}_\parallel=0$, the quantity $\delta \rho(E,\mathbf{q}_\parallel=0) \rightarrow \delta \rho(E)$ corresponds to the spatially averaged disorder-induced LDOS. Furthermore, at constant energy, the QPI pattern described by Eq.~\eqref{Eq:QPI} provides a map in reciprocal space of the possible scattering processes.

It is worth noting that the T-matrix method described here is exact, since it takes into account the summation to all orders in the impurity strength. Simpler alternative to the T-matrix is the Born approximation~\cite{balatsky2006impurity}, where only the first-order scattering is considered, as well as the joint density of states (JDOS) technique, for which the Fourier transform of the LDOS is predicted to be described by an auto-correlation of the surface spectral function~\cite{roushan2009topological}:
\begin{eqnarray}
 \label{Eq:JDOS}
 J(E,\mathbf{q}_\parallel)=\int_{BZ_\parallel} \dfrac{\mathrm{d}^2 \mathbf{k}_\parallel}{S_{BZ_\parallel}}{A}_s(E,\mathbf{k}_\parallel){A}_s(E,\mathbf{k}_\parallel+\mathbf{q}_\parallel)~.
\end{eqnarray}
However, in general, the QPI patterns based on the T-matrix and JDOS calculations do not coincide. This is because the phase factors in the Green's function matrix are not taken into account in the JDOS\cite{fang2013theory}. The QPI pattern computed with Eq.~(\ref{Eq:QPI}) is taking these interference effects into account.


\section{Results for normal UTe$_2$\label{Sec:ResultsNormal}}

We start our study by considering UTe$_2$ in its normal state, thus setting the superconducting parameter to zero: $\Delta_0=0$. The parameters used for the tight-binding model are given in Appendix~\ref{App:TBparameters}. We fix the value $\eta=0.1$~meV for the quasiparticle damping, consistent with the experimental observations in Ref.~\onlinecite{experimentalpaper}. The value of the impurity potential used to simulate hard edges is taken to be $V_0=1000$~eV, however, any value at least one order of magnitude above the bandwidth would suffice~\cite{pinon2020surface}. The value of the local impurity potential used in the QPI calculation is $U_0=0.2$~eV. The sampling $(N_{k_\#}\times N_{k_x})$ of the $\mathbf{k}$-space along the (0-11) plane is $(200\times 200)$, and we use the value $N_{k_\perp}=5000$ for integration along the direction perpendicular to the plane (0-11) to ensure convergence of the QPI numerical results.

\begin{figure}[h!]
\includegraphics[height=8cm]{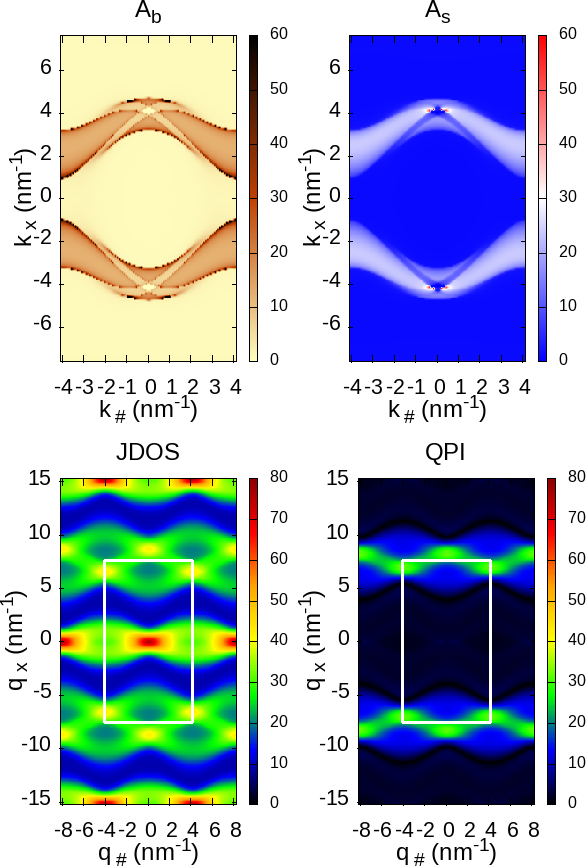}
\caption{UTe$_2$ in the normal state at $E=0$: the surface spectral functions $A_s(E,k_\#,k_x)$ and its ``bulk'' component $A_b(E,k_\#,k_x)$ (upper row), and the JDOS $J(E,q_\#,q_x)$ and QPI $\delta\rho(E,q_\#,q_x)$  (lower row). The first Brillouin zone in the (0-11) plane is marked by the white rectangle.}
\label{Fig:normalplot}
\end{figure}

Figure~\ref{Fig:normalplot} shows the bulk and surface spectral functions (upper row) and the JDOS and QPI patterns (lower row) for UTe$_2$ in its normal state at zero energy. The white rectangle corresponds to the first Brillouin zone in the plane (0-11), whose bounds are $k_x\in[-7.66,7.66]$ nm$^{-1}$ and $k_\#\in[-4.14,4.14]$ nm$^{-1}$.  We note the presence of peaks in both JDOS and QPI, that are directly related to the UTe$_2$ crystal structure. Note that in the JDOS some features are visible close to the horizontal line $q_x=0$, but they disappear in the QPI, probably due to destructive interferences.

\section{Results for superconducting UTe$_2$\label{Sec:Results}}

In this Section we study UTe$_2$ in its superconducting state for both non-chiral and chiral superconducting pairings. We first present the results for the density of states and then describe the surface spectral function and the QPI patterns. The value of the superconducting parameter is set to $\Delta_0=0.3$~meV, and the quasiparticle damping to $\eta=0.1$~meV, in agreement with the value of the thermal damping typically observed in experiments\cite{experimentalpaper}. The tight-binding parameters, given in Appendix~\ref{App:TBparameters}, and the sampling of the $\mathbf{k}$-momentum are unchanged compared with the normal case.

\subsection{Density of states}\label{Sec:DOS}

\begin{figure}[t]
\includegraphics[height=8cm]{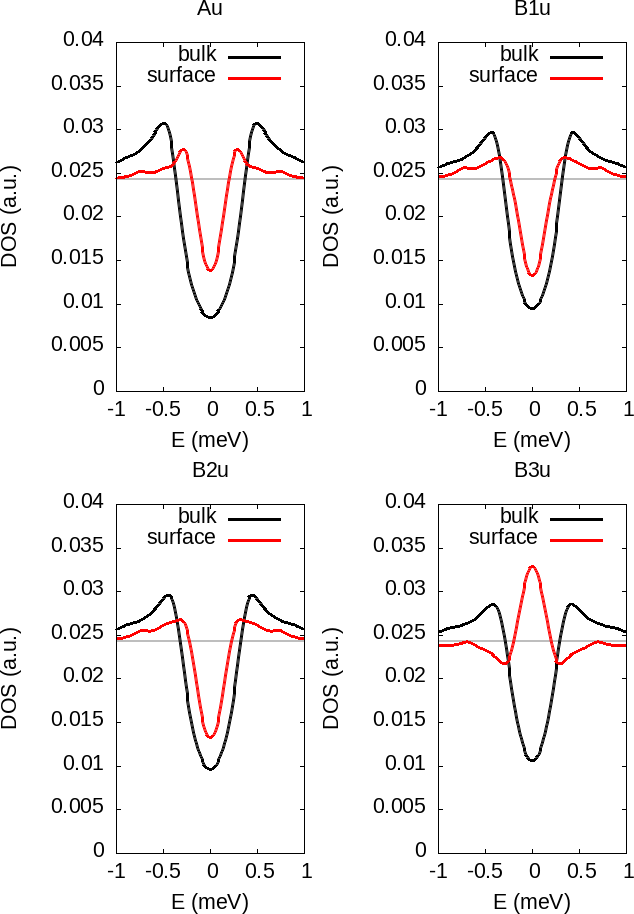}
\caption{Density of states for non-chiral pairings. The black curve represents the bulk density of states, while the red curve corresponds to the surface density of states in the presence of a (0-11) impurity plane which mimics a surface in a 3D sample. The horizontal gray line corresponds to the DOS in the normal state. At low energies, the bulk density of states for all non-chiral pairings exhibits a characteristic U-shaped profile, consistent with either gapped or nodal superconductivity in three dimensions. The $B_{3u}$ pairing is showing a pronounced zero-energy peak in the surface DOS in contrast with the other pairings still exhibiting a U-shaped profile. We have $\Delta_0=0.3$ meV and $\eta=0.1$ meV.}
\label{Fig:DOSnonchiral}
\end{figure}

\begin{figure}[t]
\includegraphics[height=8cm]{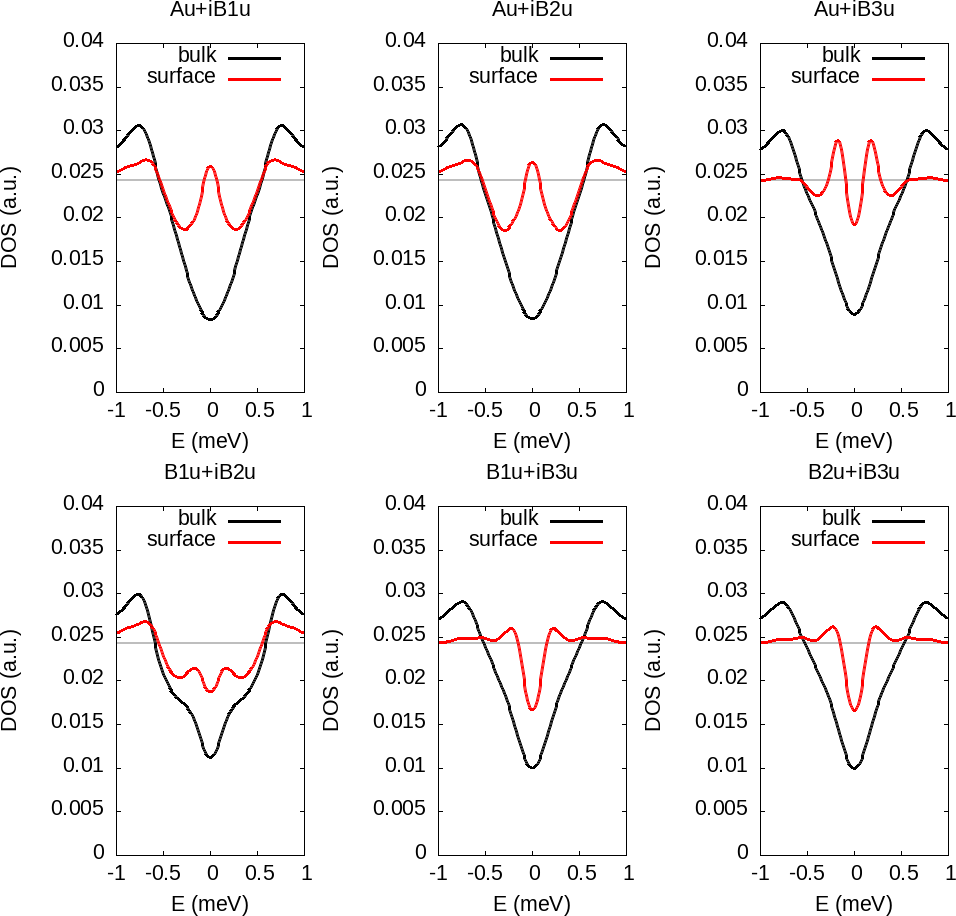}
\caption{Density of states for chiral pairings. The black curve represents the bulk density of states, while the red curve corresponds to the surface density of states in the presence of a (0-11) impurity plane which mimics a surface in a 3D sample. The horizontal gray line corresponds to the DOS in the normal state. At low energies, the bulk density of states for all chiral pairings exhibits a characteristic U-shaped profile, consistent with either gapped or nodal superconductivity in three dimensions. We take $\Delta_0=0.3$ meV and $\eta=0.1$ meV.}
\label{Fig:DOSchiral}
\end{figure}

The bulk and surface density of states, $\rho_0(E)$ and $\rho_s(E)$, are respectively computed using
\begin{eqnarray}
\rho_0(E)&=&-\dfrac{1}{\pi}\int_{BZ}\text{Im}\{\text{Tr}\big[G_\mathrm{BdG}(E,\mathbf{k})\big]\}\mathrm{d}^3\mathbf{k}
\label{rho0}~,\\
\rho_s(E)&=&-\dfrac{1}{\pi}\int_{BZ_\parallel}\text{Im}\{\text{Tr}\big[G_s(E,\mathbf{k}_\parallel)\big]\}\mathrm{d}^2\mathbf{k}_\parallel~,
\end{eqnarray}
where $G_\mathrm{BdG}(E,\mathbf{k})$ and $G_s(E,\mathbf{k}_\parallel)$ are the Bogoliubov-de-Gennes and respectively the surface Green's functions given by Eqs.~(\ref{Eq:BulkGreenFunction}) and (\ref{Eq:GS_Equation}). The integral in Eq.~\ref{rho0} is performed over the entire 3D BZ. The results are presented in Figs.~\ref{Fig:DOSnonchiral} and \ref{Fig:DOSchiral} for all the pairing symmetries considered in this study. All chiral and non-chiral pairings have a U-shaped bulk density of states at low energies, consistent with point nodes~\cite{ishihara2023chiral}. Some pairings acquire a V-shape at larger energies when other bands start contributing.

We note also that the $B_{3u}$, $A_u+iB_{1u}$ and $A_u+iB_{2u}$ pairing symmetries exhibit a maximal value at zero energy in the (0-11) surface density of states, indicating the formation of surface states in the vicinity of zero energy, while the other pairings have a minimal value at zero energy and two maxima at subgap energies of the order of $E\approx \pm 0.2$ meV, indicating the formation of surface states close to this energy value. The DOS profile can be affected by additional allowed symmetry d-vector contributions, as discussed in Appendix~\ref{App:DOS_dvector}. In particular,  the zero-energy peak observed for the $B_{3u}$ order parameter can split upon adding additional d-vector terms.

\subsection{Bulk spectral function\label{App:Bulk_SF}}

\begin{figure}[t]
\includegraphics[height=8cm]{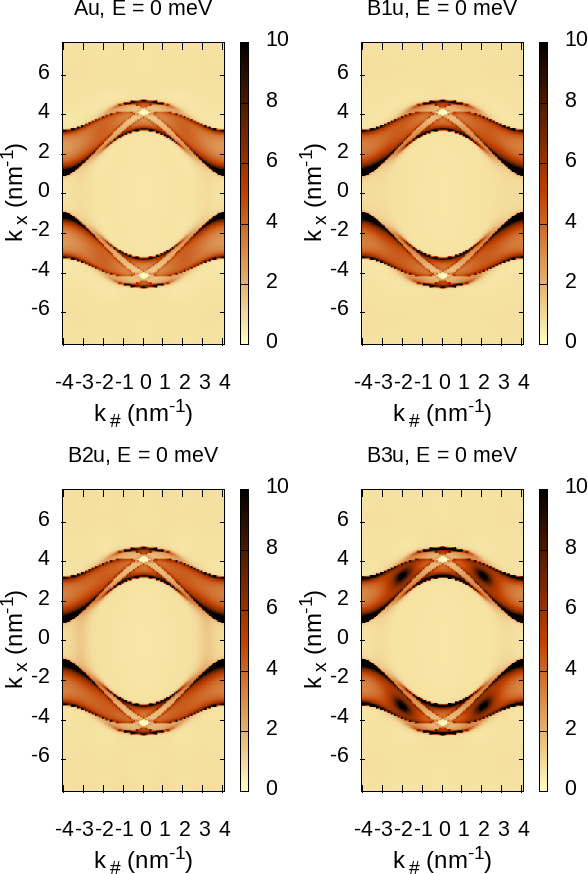}
\caption{Bulk spectral function in the (0-11)-plane  for non-chiral pairings at $E=0$. We take $\Delta_0=0.3$ meV and $\eta=0.1$ meV. The pairings exhibit similar spectral functions, with differences arising from their distinct nodal structures.}
\label{Fig:A0_nonchiral}
\end{figure}

\begin{figure}[t]
\includegraphics[height=8cm]{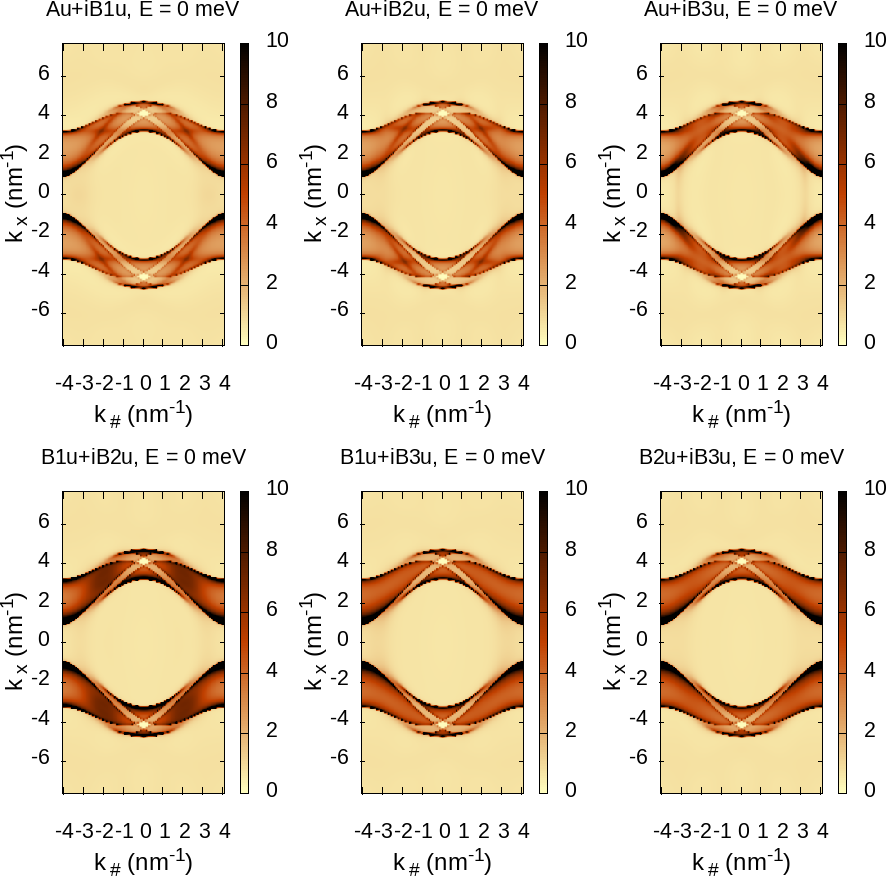}
\caption{Bulk spectral function in the (0-11)-plane  for chiral pairings at $E=0$. We take $\Delta_0=0.3$ meV and $\eta=0.1$ meV.}
\label{Fig:A0_chiral}
\end{figure}

The bulk contribution to the spectral function $A_b(E,k_\#,k_x)$, defined by Eq.~\ref{Eq:BulkSpectral}, essentially represents the direct projection of the imaginary part of the bulk Green function onto the considered surface, i.e. (0-11)- plane. Figures~\ref{Fig:A0_nonchiral} and~~ \ref{Fig:A0_chiral} show this quantity at zero-energy, for both non-chiral and chiral pairings. Note that the overall amplitude for $A_b(E,k_\#,k_x)$ is reduced when UTe$_2$ is in its superconducting state compared with UTe$_2$ in its normal state. Apart from this amplitude reduction, we observe that the bulk spectral function profile as a function $k_\#$ and $k_x$ remains generally similar to the normal state one (see Fig.~\ref{Fig:normalplot}). However, as discussed in detail in Ref. ~\onlinecite{experimentalpaper}, for some superconducting order parameters, in particular $B_{2u}$ and $B_{3u}$ among the non-chiral order parameters, one expects the presence of extra features, in particular of nodal points. Here, some of these nodes are masked by significant quasiparticle damping, chosen to match the realistic experimental conditions. However, they can become visible for significantly smaller damping, as illustrated in Fig.~\ref{Fig:Nodal_Bulk} in Appendix~\ref{App:Node_Structure}.

As explained in Appendix~\ref{App:bulk}, the symmetries of the BdG Hamiltonian imply that the bulk contribution to the spectral function is mirror-symmetric along the $k_x$ and $k_\#$ axis for any pairings, in agreement with what Figs.~\ref{Fig:A0_nonchiral} and \ref{Fig:A0_chiral} show. It relies on mirror inversion $m_x$ and 2-fold rotation for the non-chiral pairings and for $A_u+iB_{3u}$ and $B_{1u}+iB_{2u}$ pairings. For $A_u+iB_{1u}$, $A_u+iB_{2u}$, $B_{1u}+iB_{3u}$ and $B_{2u}+iB_{3u}$, the magnetic symmetries $\mathcal{T}m_x$ and $\mathcal{T}C_{2x}$ protect the mirror symmetries of the bulk contribution.

\subsection{Surface spectral function}\label{Sec:Spec}

\begin{figure}[t]
\includegraphics[height=8cm]{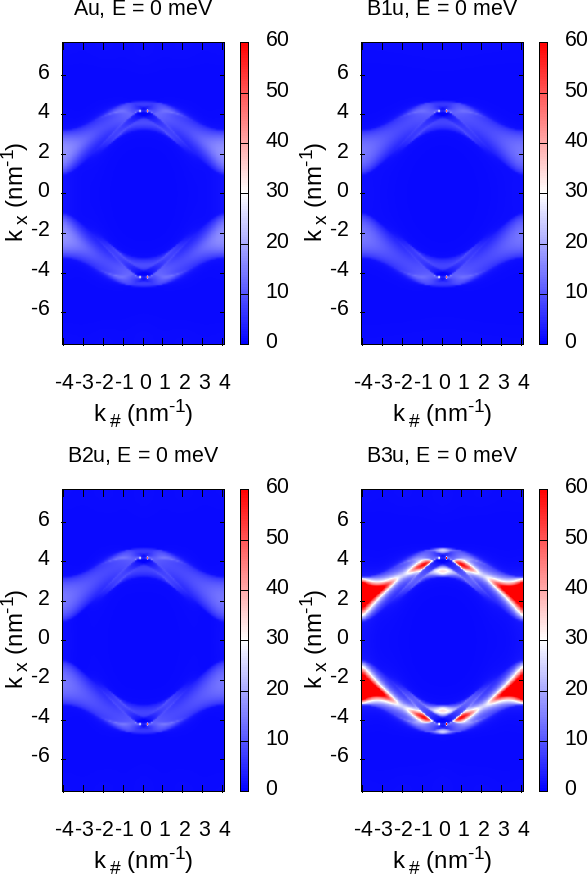}
\caption{Surface spectral function as a function of $k_\#$ and $k_x$ at $E=0$ for non-chiral pairings. The gap value is fixed at $\Delta_0=0.3$ meV and the quasiparticle damping at $\eta=0.1$ meV.}
\label{Fig:ASurface_E0_nonchiral} 
\end{figure}

\begin{figure}[t]
\includegraphics[height=8cm]{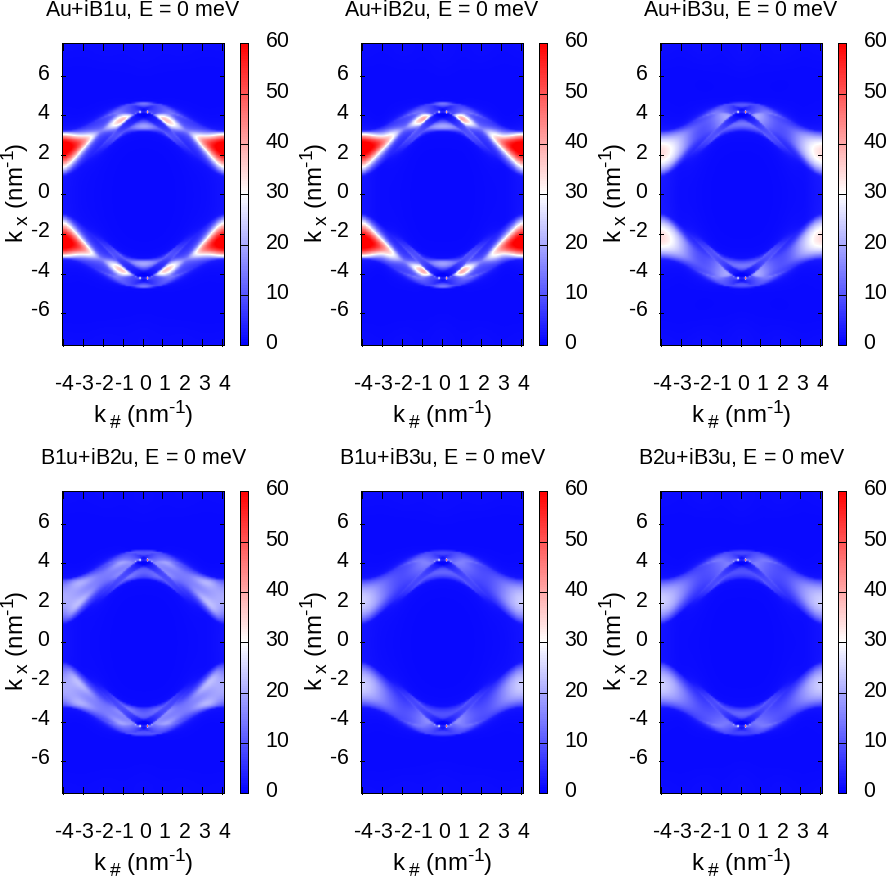}
\caption{Surface spectral function as a function of $k_\#$ and $k_x$ at $E=0$ for chiral pairings. The gap value is fixed at $\Delta_0=0.3$ meV and the quasiparticle damping at $\eta=0.1$ meV.}
\label{Fig:ASurface_E0_chiral}
\end{figure}

The surface spectral function $A_s(E,\mathbf{k}_\parallel)$ is calculated from Eq.~(\ref{Eq:SurfaceSpectral}), and plotted at zero energy in Figs.~\ref{Fig:ASurface_E0_nonchiral} and~\ref{Fig:ASurface_E0_chiral} as a function of $k_\#$ and $k_x$ in the first Brillouin zone. For non-chiral pairings, we observe similar profiles of the surface spectral function for $A_u$, $B_{1u}$ and $B_{2u}$, with an amplitude which is of the same order compared with that of the normal surface spectral function displayed in Fig.~\ref{Fig:normalplot}. On the contrary, for the $B_{3u}$ pairing, we obtain a strong enhancement in the amplitude of the surface spectral function compared with the normal one and compared with the other non-chiral pairings. This is the signature of the emergence of surface states at zero energy in the $B_{3u}$ case. Figure~\ref{Fig:ASurface_E0_chiral} shows that the amplitude is also enhanced for $A_u+iB_{1u}$ and  $A_u+iB_{2u}$, compared with the normal state, whereas it stays quite similar to the normal state for the other chiral pairings. Thus, both the density of states and the surface spectral function show the emergence of surface superconducting states for $B_{3u}$, $A_u+iB_{1u}$ and $A_u+iB_{2u}$ pairings close to zero-energy. In general, topological surface states form as arcs between the projection of the bulk nodes to the surface. Among the non-chiral OPs for which we have investigated the nodal structure, this seems to be the case only for $B_{3u}$. For the other pairings, the surface states are not connected to the nodal structure, and they form at finite energy; their origin is thus unclear.

\begin{figure}[t]
\includegraphics[height=8cm]{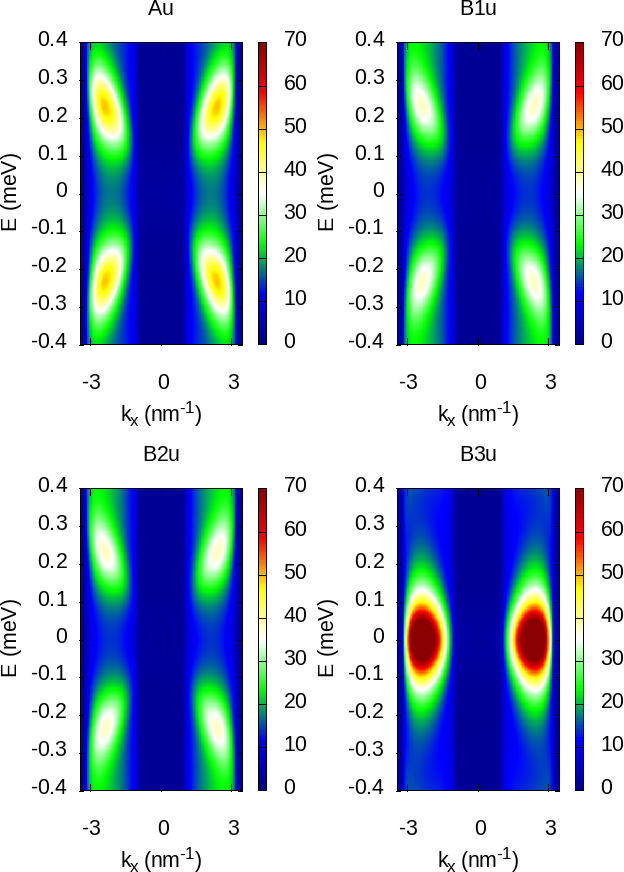}
\caption{Surface band dispersion along $k_x$ for non-chiral pairings at $k_\#=-4.14$ nm$^{-1}$, $\Delta_0=0.3$~meV and $\eta=0.1$~meV.}
\label{Fig:E_nonchiral}
\end{figure}

\begin{figure}[t]
\includegraphics[height=8cm]{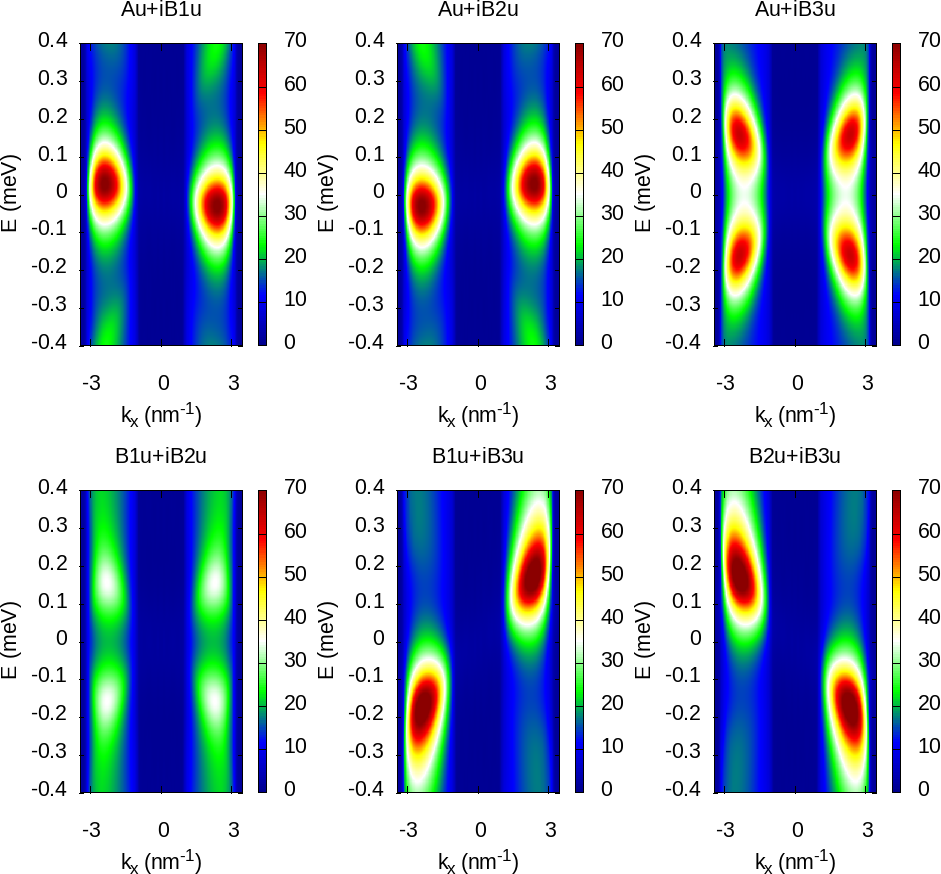}
\caption{Same as in Fig.~\ref{Fig:E_nonchiral} for chiral pairings.}
\label{Fig:E_chiral}
\end{figure}

To understand this in more detail, in Figs~\ref{Fig:E_nonchiral} and \ref{Fig:E_chiral} we plot the surface band dispersion as a function of~$k_x$ for $k_\#=-4.14$ nm$^{-1}$, i.e. along one of the edges of the BZ$_\parallel$, and along a close path ABCDA of the BZ$_\parallel$ in Appendix~\ref{App:band_ABCD}. This clearly shows the emergence of sub-gap surface states centered at zero-energy for a $B_{3u}$ pairing, and very close to zero-energy for $A_u+iB_{1u}$ and $A_u+iB_{2u}$ parings. All other pairings exhibit subgap surface states centered at a finite energy of $\approx \pm 0.2$ meV.
These results are summarized in Table~\ref{Table:Summary_Results}.  The complete surface spectral function dispersion for the entire Brillouin zone is given in Figs.~\ref{Fig:ASurface_E_nonchiral} and \ref{Fig:ASurface_E_chiral} in Appendix \ref{App:A_E}, and is consistent with these observations. The observed surface states could potentially be topological surface states~\cite{sato2017topological}. However, as this aspect is beyond the scope of the present work, we refer to them simply as surface states.

The choice of damping value $\eta$ is constrained by experiments~\cite{experimentalpaper}. By decreasing the value of $\eta$ in our simulation, the difference in amplitude between the superconducting and normal cases is accentuated, and the nodes presented schematically in Ref.~\cite{experimentalpaper} for $B_{2u}$ and $B_{3u}$ become visible in the surface spectral function (not shown here, but presented in the Methods section of Ref.~\cite{experimentalpaper}). We have checked that the surface states associated with these nodes do not contribute to the QPI scattering, and that the conclusions of the present analysis are unchanged by the value of $\eta$, and by the nodal point visibility.

The symmetry constraints imposed by the crystalline symmetries and the time reversal on the edge contribution to the surface state are explained in detail in Appendix~\ref{App:Symmetry_Edge}. The non-chiral pairings are mirror symmetric along both $k_x$ and $k_\#$ mirror axis, as in the bulk case, as shown in Fig.~\ref{Fig:ASurface_E0_nonchiral}. These constraints arise from the $m_x$ and $\mathcal{T}m_x$ symmetries. For chiral pairings, we differentiate between those that preserve $m_x$ symmetry and those that do not, similar to the bulk case. The surface spectral function of the order parameter $A_u+iB_{3u}$ and $B_{1u}+iB_{2u}$ is mirror-symmetric along the $k_x$ axis and protected by the $m_x$ symmetry. The surface spectral function for the other order parameters that break $m_x$ remains mirror symmetric along the $k_\#$ axis, a symmetry protected by $\mathcal{T}m_x$. These symmetry arguments hold for an arbitrary energy $E$ (see Figs.~\ref{Fig:ASurface_E_nonchiral} and \ref{Fig:ASurface_E_chiral} in Appendix \ref{App:A_E} for the plots at finite energy). At zero energy, the action of particle-hole symmetry leads to an enlarged symmetry group, making all superconducting order parameters symmetric along both the $k_x$ and $k_\#$ axis, as shown in Figs.~\ref{Fig:ASurface_E0_nonchiral} and~\ref{Fig:ASurface_E0_chiral}.

\subsection{Spin-resolved surface spectral function}\label{Sec:Spinresolved}

\begin{figure}[t]
\includegraphics[height=4cm]{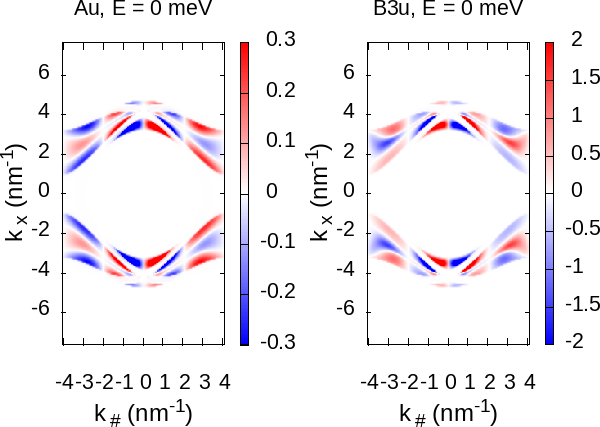}
\caption{Spin-resolved surface spectral function along $\widehat{x}$-axis for $A_{u}$ and $B_{3u}$ pairings at  $E=0$, $\Delta_0=0.3$~meV and $\eta=0.1$~meV. Note that the spin-resolved surface spectral function along $\widehat{x}$-axis is equal to zero for $B_{1u}$ and $B_{2u}$ pairings. The symmetry of the spin texture can be understood from the structure of the superconducting order parameter for each pairing, as detailed in Appendix~\ref{App:SpinPolarization}.}
\label{Fig:ASurface_E0_Sx_nonchiral}
\end{figure}

\begin{figure}[t]
\includegraphics[height=8cm]{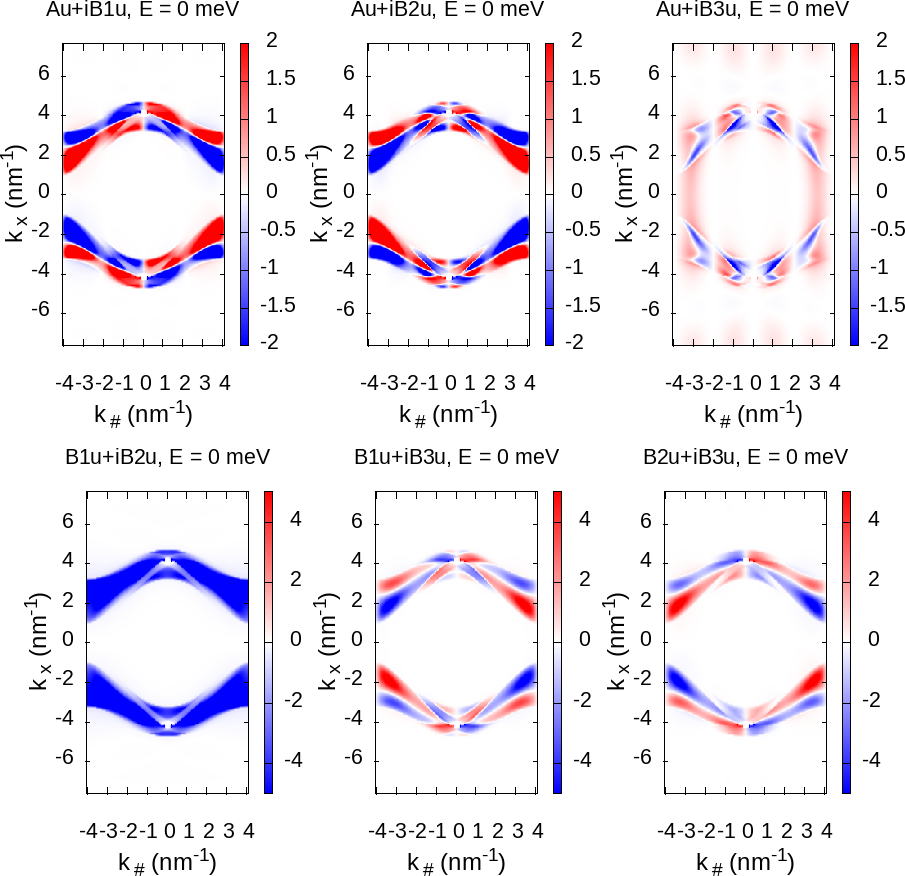}
\caption{Spin-resolved surface spectral function along $\widehat{x}$-axis for chiral pairings at  $E=0$, $\Delta_0=0.3$~meV and $\eta=0.1$~meV. Since chiral pairings have superconducting order parameter components along all spin directions, each chiral pairing exhibits non-zero spin polarization in every spin channel.}
\label{Fig:ASurface_E0_Sx_chiral}
\end{figure}

The tight-binding model used in this work, described in Eq.~(\ref{Eq:TB_model}), is spin $\uparrow, \downarrow$ symmetric, with the two spin components completely decoupled. This means that any spin polarization of the spectral function comes from the triplet superconducting order parameter defined in Eq.~(\ref{Eq:SC_OP}). Figures~\ref{Fig:ASurface_E0_Sx_nonchiral} and \ref{Fig:ASurface_E0_Sx_chiral} display the spin-resolved spectral function projected along the $\widehat{x}$-axis for both chiral and non-chiral pairings. In the non-chiral case, we plot this quantity only for $A_{u}$ and $B_{3u}$ pairings because it cancels for $B_{1u}$ and $B_{2u}$ pairings. For the sake of simplicity, we present the spin-resolved spectral functions along the $\widehat{y}$-axis and $\widehat{z}$-axis in Appendix~\ref{App:SpinResolved}.

Since the normal state Hamiltonian is spin-independent, the bulk contribution results in a vanishing spin polarization for all pairings. On the contrary, the surface contribution can acquire a non-trivial spin-polarization depending on the form of the superconducting order parameter. From Figs.~\ref{Fig:ASurface_E0_Sx_nonchiral},~\ref{Fig:ASurface_E0_Sy_nonchiral} and~\ref{Fig:ASurface_E0_Sz_nonchiral}, we see that the surface spectral function is polarized along $\widehat{x}$-axis for $B_{3u}$, along $\widehat{y}$-axis for $B_{2u}$ and  along $\widehat{z}$-axis for $B_{1u}$, while the other pairings exhibit a more isotropic spin response (see Figs.~\ref{Fig:ASurface_E0_Sx_chiral}, ~\ref{Fig:ASurface_E0_Sy_chiral} and~\ref{Fig:ASurface_E0_Sz_chiral}). In consequence, the $B_{3u}$ pairing case is the only one which has a spin-resolved surface spectral function with a polarization parallel to the $\widehat{x}$-axis, i.e. parallel to the (0-11) plane.

\subsection{QPI patterns}\label{Sec:QPI}

\begin{figure}[t]
\includegraphics[height=8cm]{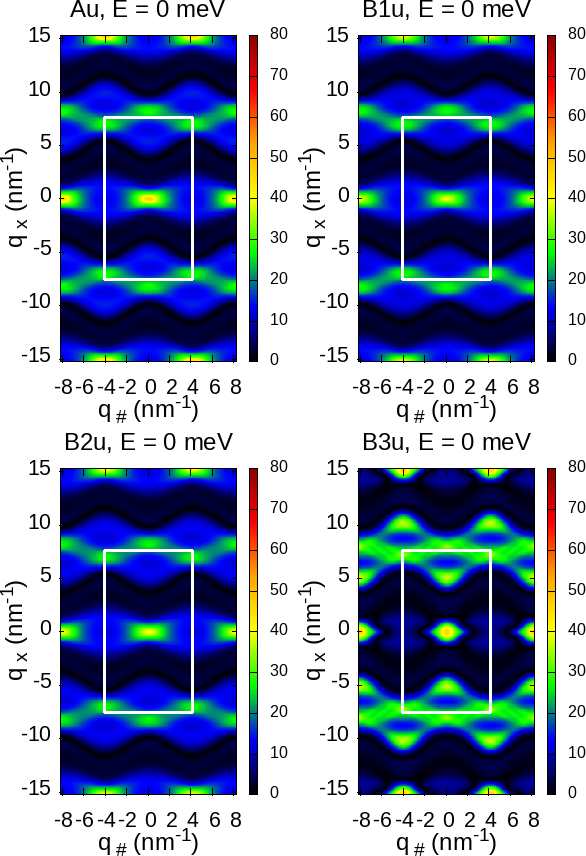}
\caption{QPI patterns as a function of $q_\#$ and $q_x$ for non-chiral pairings at $E=0$, $U_0=0.2$ eV, $\Delta_0=0.3$~meV and $\eta=0.1$~meV. Because of their similar surface spectral functions, the $A_u$, $B_{1u}$ and $B_{2u}$ pairings exhibit relatively similar QPI patterns. In contrast, the $B_{3u}$ pairing displays an additional peak from from surface states scattering. The distinct nodal structures of the order parameters are not reflected in the QPI patterns due to the strong quasiparticle damping $\eta$.}
\label{Fig:QPI_E0_nonchiral}
\end{figure}

\begin{figure}[t]
\includegraphics[height=8cm]{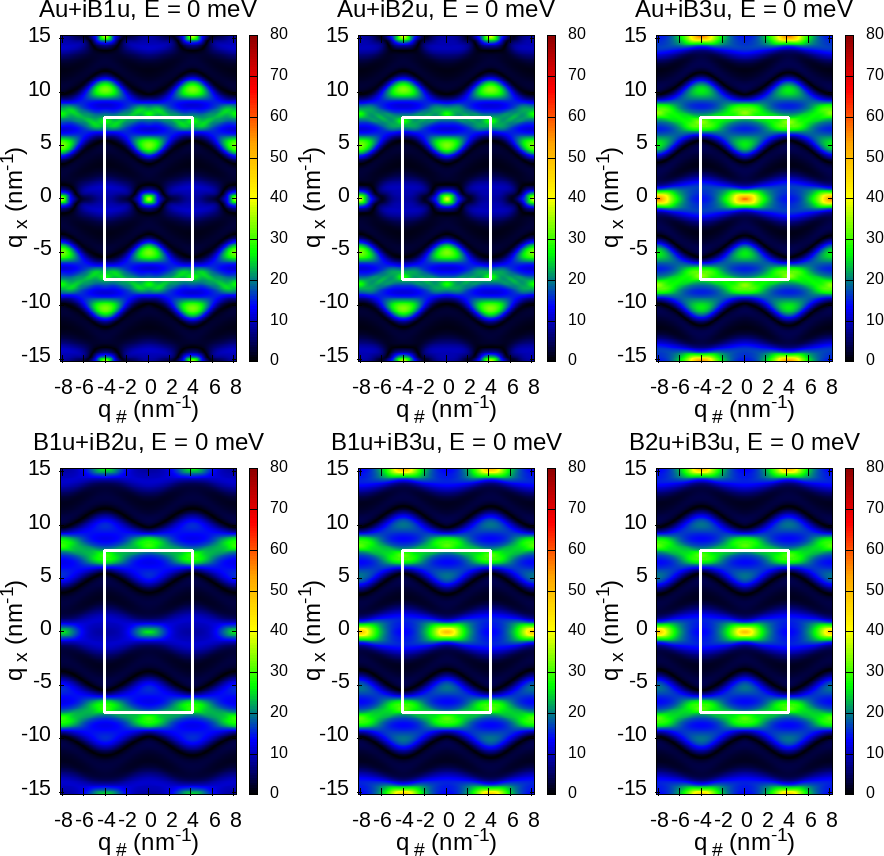}
\caption{QPI patterns as a function of $q_\#$ and $q_x$ for chiral pairings at $E=0$, $U_0=0.2$ eV, $\Delta_0=0.3$~meV and $\eta=0.1$~meV. Similar to $B_{3u}$ pairing, $A_u+iB_{1u}$ and $A_u+iB_{2u}$ display an additional peak from from surface states scattering. This peak is present but with weaker intensity for $A_u+iB_{3u}$. The distinct nodal structures of the order parameters are not reflected in the QPI patterns due to the strong quasiparticle damping $\eta$.}
\label{Fig:QPI_E0_chiral}
\end{figure}

Figures~\ref{Fig:QPI_E0_nonchiral} and \ref{Fig:QPI_E0_chiral} show the QPI patterns for the superconducting state of UTe$_2$ described by various non-chiral and chiral OPs at zero energy. Once more we consider an impurity localized on the (0-11)-plane with impurity potential $U_0=0.2$ eV. There are two key differences compared with the QPI pattern obtained for UTe$_2$ in the normal state (see Fig.~\ref{Fig:normalplot}): (i)~the presence of a finite intensity along a horizontal band close to $q_x=0$ for all the pairings, which was absent in the normal state, and (ii)~the presence  at $q_\#=0$ and finite  $q_x$ of an additional peak, denoted ${\bf q}_1$ for some of the OPs, especially $B_{3u}$, $A_u+iB_{1u}$, $A_u+iB_{2u}$, $A_u+iB_{3u}$, $B_{1u}+iB_{3u}$ and $B_{2u}+iB_{3u}$.

\begin{figure}[h!]
\includegraphics[width=8.cm]{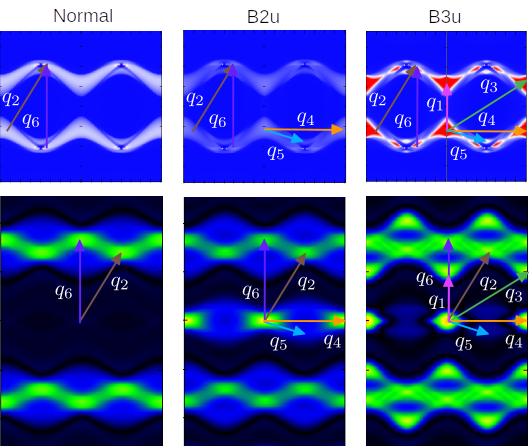}
\caption{Scattering vectors ${\bf q}_i$, with $i\in[1,6]$, for normal state, $B_{2u}$ pairing and $B_{3u}$ pairing at $E=0$. The top plots are the surface spectral functions while the bottom plots are the QPI patterns. The $(k_\#,k_x)$-dependence of the surface spectral function allows us to understand the origin and the position of the peaks in the QPI patterns.}
\label{Fig:ScatteringVector}
\end{figure}

The various peaks in the QPI patterns are identified by the vectors ${\bf q}_1$ to ${\bf q}_6$. To help understand the presence of these peaks, we have marked the scattering vectors associated to each of these peaks in Fig.~\ref{Fig:ScatteringVector}, for both normal UTe$_2$ and superconducting UTe$_2$ with pairings $B_{2u}$ and $B_{3u}$. In the normal state case, there are only two peaks, characterized by the scattering vectors ${\bf q}_2$ and ${\bf q}_6$ that are related to the crystal structure. For the $B_{2u}$ pairing two additional peaks are visible,  characterized by vectors ${\bf q}_4$ and~${\bf q}_5$. Finally, for the $B_{3u}$ pairing, two other peaks are visible, characterized by vectors ${\bf q}_1$ and~${\bf q}_3$. Thus, at $E=0$, there are two peaks in the normal state, four peaks for $B_{2u}$, and six peaks for $B_{3u}$. The other non-chiral and chiral pairings show similar structure to either $B_{2u}$ and $B_{3u}$. We have identified the correspondence between the scattering vectors and the regions of high intensity in the surface spectral function. We note, however, that while some peaks would be expected from the spectral function analysis, their absence is related to destructive interference, leading to forbidden scattering processes between some of the bands.

It should also be emphasized that the QPI patterns depend slightly on the value of the impurity amplitude $U_0$ (see Eqs.~(\ref{Eq:Tmatrix}) to~(\ref{Eq:LDOS})). However, we have checked that the main features and the conclusion of our analysis do not change when $U_0$ varies.

The evolution of the QPI pattern with increasing energy $E$ from zero up to the value 0.25 meV is given in Figs.~\ref{Fig:QPInonchiral_ne0} and \ref{Fig:QPIchiral_ne0} in Appendix \ref{App:E_QPI}. 
These figures indicate that the intensity of the QPI patterns reflects the intensity of the spectral function, and this is especially true for the $q_1$ peak whose maximum of intensity as a function of energy follows the maximum intensity in Figs~\ref{Fig:E_nonchiral} and \ref{Fig:E_chiral} for each OP, as noted in Table~\ref{Table:Summary_Results}. 

\begin{figure}[h!]
\includegraphics[height=4.5cm]{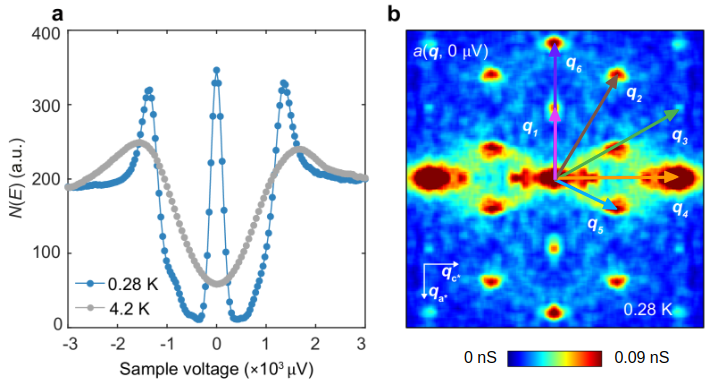}
\caption{Experimental measurement of differential  conductance for (left panel) UTe$_2$ in the normal state (4.2 K) and Andreev differential conductance in the superconducting state (0.28 K), and (right panel) QPI pattern at zero energy\cite{experimentalpaper}.}
\label{Fig:expDOS}
\end{figure}

\begin{figure}[h!]
\includegraphics[width=9.cm]{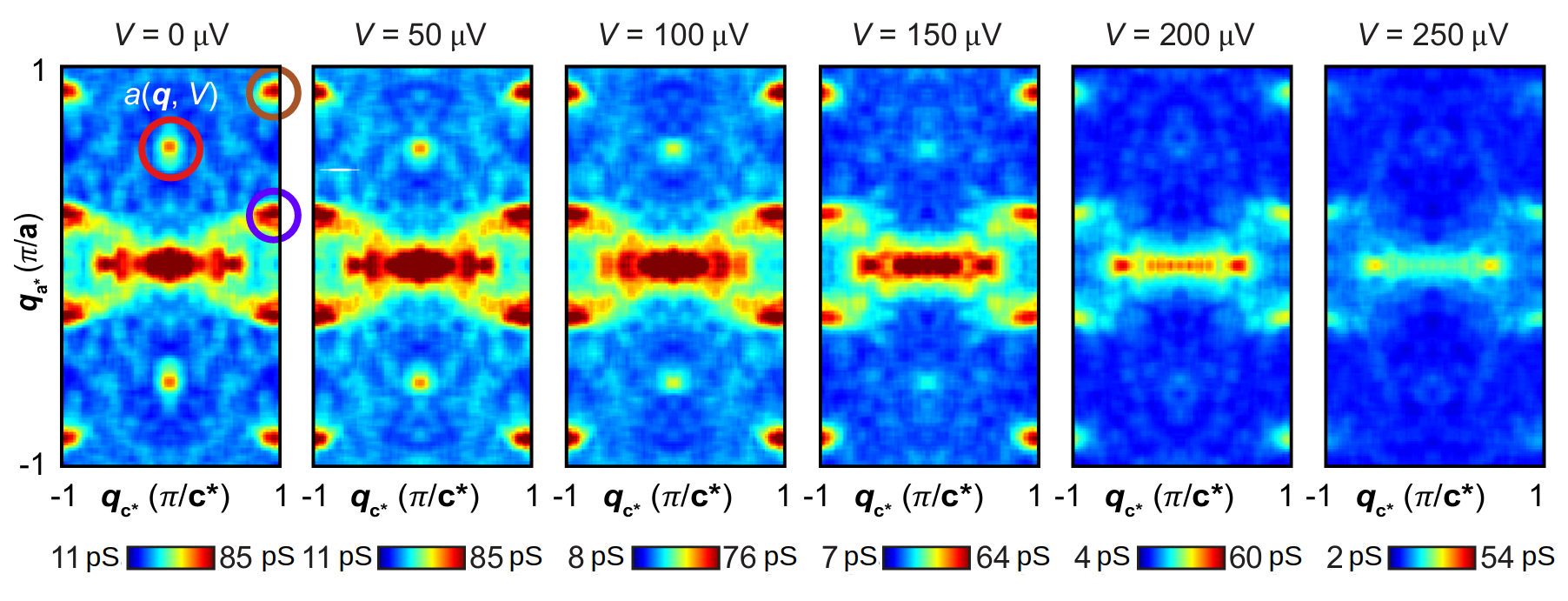}
\caption{Experimental results for QPI pattern plotted in the first Brillouin zone, at increasing energy $E\in\{0,50,100,150,200,250\}$ $\mu$eV\cite{experimentalpaper}.}
\label{Fig:expQPI}
\end{figure}

We compare these results with the experimental observations in Ref.~\onlinecite{experimentalpaper} presented in Figs.~\ref{Fig:expDOS} and \ref{Fig:expQPI}. In particular, the experimental results show a peak in the DOS at zero energy, as well as a non-dispersing ${\bf q}_1$ feature. Among the non-chiral states, that are most likely to describe the physics of UTe$_2$, the $B_{3u}$ OP is the only one that has a ${\bf q}_1$ peak that is stable in position and remains visible up to $E\approx 0.15$ meV. This observation, combined with the fact that $B_{3u}$ is the only OP among the non-chiral one that exhibits a peak at zero energy in the DOS, indicates that $B_{3u}$ is the most likely candidate to describe the superconducting order parameter of UTe$_2$.

\subsection{Summary of the results}

Table~\ref{Table:Summary_Results} gives a short summary of the results obtained in this section.
\begin{table}[h!]
\centering
\begin{tabular}{|c|c|c|c|}
\hline
Pairing & DOS max & ${\bf q}_1$ peak in QPI at $E=0$  \\
\hline
\hline
$A_u$ &  $E\approx \pm 0.24$  & -- \\
\hline
$B_{1u}$ &  $E\approx \pm 0.24$   & -- \\
\hline
$B_{2u}$ & $E\approx \pm 0.24$   & -- \\
\hline
$B_{3u}$ &  $E=0$ & Yes, strong amplitude \\
\hline
\hline
$A_u+iB_{1u}$ &   $E=\pm0.03$  &  Yes, strong amplitude \\
\hline
$A_u+iB_{2u}$ &  $E=\pm0.03$  & Yes, strong amplitude  \\
\hline
$A_u+iB_{3u}$ & $E\approx \pm 0.18$  &  Yes, moderate amplitude \\
\hline
$B_{1u}+iB_{2u}$ &  $E\approx \pm 0.18$  & -- \\
\hline
$B_{1u}+iB_{3u}$ &  $E\approx \pm 0.2$   & Yes, weak amplitude \\
\hline
$B_{2u}+iB_{3u}$ &   $E\approx \pm 0.22$ &  Yes, weak amplitude  \\
\hline
\end{tabular}
\caption{Summary of the results for both chiral and non-chiral pairings indicating in the second column the position of the DOS maximum, consistent with the maximum in the spectral function dispersion, and in the third column the presence of a ${\bf q}_1$ peak in the QPI at $E=0$.  The values of $E$ are given in meV. We take  $\Delta_0=0.3$~meV, $\eta=0.1$~meV and $U_0=0.2$~eV.}
\label{Table:Summary_Results}
\end{table}


\section{Discussions and conclusion\label{Sec:Discussions}}

As outlined in the introduction, determining the pairing symmetry of UTe$_2$ is challenging and requires a multi-method approach.  Experimental data available concerning time-reversal symmetry breaking seems to indicate that a non-chiral pairing is the most probable hypothesis~\cite{gu2023detection,theuss2024single}. Currently, there is active debate on whether the pairing symmetry is $B_{2u}$ or $B_{3u}$. Our analysis can help to resolve this debate by providing clear predictions to distinguish especially between the non-chiral pairing symmetries $A_u$, $B_{1u}$, $B_{2u}$, and $B_{3u}$ through STM experiments.  Indeed, by comparing our results to the experimental ones\cite{experimentalpaper}, we can give strong arguments that the pairing state of superconducting UTe$_2$ is $B_{3u}$. This is based notably on the existence of the ${\bf q}_1$ peak in the QPI patterns and on the presence of the zero-energy peak in the surface DOS.

If one also takes into account the chiral OPs, the $A_u+iB_{1u}$, $A_u+iB_{2u}$, $A_u+iB_{3u}$, $B_{1u}+iB_{3u}$ or $B_{2u}+iB_{3u}$ also show a ${\bf q}_1$  feature. However, the latter three can be eliminated from the possible scenarios, since they have minimum in its DOS at zero-energy which contradicts the experimental results\cite{experimentalpaper} as shown in the left panel of Fig.~\ref{Fig:expDOS}. The discrimination between the remaining pairings, $B_{3u}$, $A_u+iB_{1u}$ and $A_u+iB_{2u}$, could be done using various other methods.  As noted earlier, the chiral order parameters appear highly unlikely, a conclusion reinforced by the splitting of the zero-energy Andreev peak~\cite{gu2025pair} under proximity with a $s$-wave superconductor, which suggests a time-reversal symmetric superconducting state. In the future spin-resolved experiments may also be used to distinguish between these pairings, since we have shown in Section~\ref{Sec:Spinresolved} that they have different spin-resolved spectral functions.

{\it Note added.} After our work was completed, we became aware of Ref.~\cite{christiansen2025nodal} which reports a model for the band structure and the superconductive topological surface states of UTe$_2$ using a different technique.

\begin{acknowledgements}
A.C. thanks the CALMIP supercomputing center for the allocation of HPC numerical resources through Project M23023 supported by a French government grant managed by the Agence Nationale de la Recherche under the Investissements d'avenir program (ANR-21-ESRE-0051). J.C.S.D. acknowledges support from the Royal Society under Award R64897. J.P.C., K.Z. and J.C.S.D. acknowledge support from Science Foundation Ireland under Award SFI 17/RP/5445. S.W. and J.C.S.D. acknowledge support from the European Research Council (ERC) under Award DLV-788932. Q.G., K.Z., J.P.C., S.W., and J.C.S.D. acknowledge support from the Moore Foundation’s EPiQS Initiative through Grant GBMF9457. K.Z. acknowledges funding from the European Union’s Horizon Europe research and innovation programme under the Marie Skłodowska-Curie Actions Postdoctoral Global Fellowship “QUASAR”, grant agreement No. 101203931.
\end{acknowledgements}


\appendix



\section{Additional information on model and methods}\label{App:ModelMethod}

\subsection{Tight-binding parameters}\label{App:TBparameters}

\begin{figure}[h!]
\includegraphics[width=9.cm]{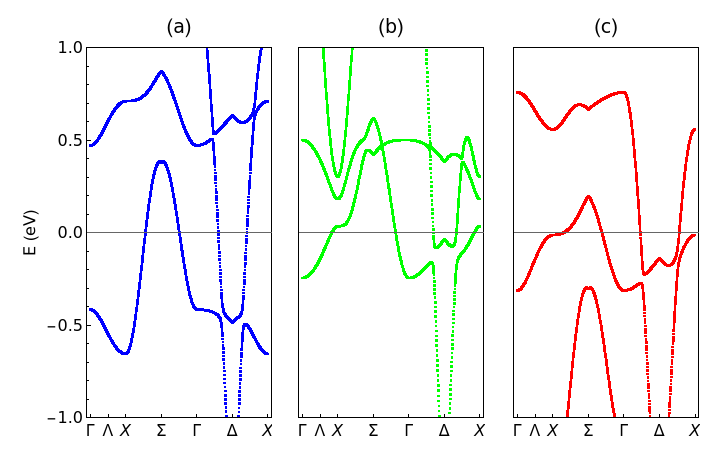}
\caption{Energy band plotted along $\Gamma\Lambda$X$\Sigma\Gamma\Delta$X path in the 3D Brillouin zone for (a) DFT parameters of Ref.~\onlinecite{theuss2024single}, (b) QO parameters of Ref.~\onlinecite{theuss2024single}, and (c) this work (see Tab.~\ref{Table:parameter_set}).}
\label{Fig:band_appendixA}
\end{figure}

To describe the band structure of UTe$_2$, we use a 4-orbital tight-binding model introduced by Theuss et al.~\cite{theuss2024single}, and parameter values modified compared with those obtained from DFT or QO calculations. The parameters entering in Eq.~(\ref{Eq:TB_model}) are given in the last column of Tab.~\ref{Table:parameter_set}. It leads to an area for the electron and hole pockets of $\mathcal{A}_e\approx\mathcal{A}_h\approx 34.9$ nm$^{-2}$ at $k_c=0$, a value which is close to the value ($\mathcal{A}_e\approx\mathcal{A}_h\approx \ 33.6$ nm$^{-2}$) measured in recent experiments~\cite{eaton2024quasi,weinberger2024quantum}. As shown in Fig.~\ref{Fig:band_appendixA}, the topology of the band structure close to energy equal to zero is qualitatively similar, with two bands crossing the Fermi level, one with a maximum close to the $\Sigma$ point, and one with a minimum close to the $\Delta$ point, giving rise respectively to the electron and hole pockets consistent with the quantum oscillation measurements~\cite{eaton2024quasi,weinberger2024quantum}. We also note that, for our parameter set, the band exhibits a flattening near zero energy along the $\Sigma X$ direction close to the $X$ point, in a manner quite similar to what is observed for the QO parameter set of Ref.~\cite{theuss2024single}. The exact form of the bands far away from the Fermi level does not affect the physics since we focus on SC gaps of the order of sub-mV.

\begin{table}[h!]
\begin{tabular}{|c||c|c|c|}
\hline
 Parameter&DFT~\cite{theuss2024single}&QO~\cite{theuss2024single}&This work\\
 \hline
 \hline
 $\mu_U$ &0.35 & 0.35 & -0.355 \\
 \hline
 $\Delta_U$ & 0.4 & 0.05 &  0.38 \\
 \hline
 $t_U$ &0.15 & 0.1 & 0.17 \\
 \hline
 $t'_U$ & 0.08 & 0.08 & 0.08 \\
 \hline
 $t_{ch,U}$ &0.01 & 0.01 & 0.015 \\
 \hline
 $t'_{ch,U}$ &0 & 0 &  0.01 \\
 \hline
 $t_{z,U}$ &-0.03 & 0.04 & -0.0375 \\
 \hline
 $\mu_{Te}$ & -1.8 & -1.8  &  -2.25\\
 \hline
 $\Delta_{Te}$ & -1.5 & -1.5 & -1.4\\
 \hline
 $t_{Te}$ & -1.5 & -1.5 & -1.5  \\
 \hline
 $t_{ch,Te}$ & 0 & -0.03 &  0\\
 \hline
 $t_{z,Te}$ & -0.05 & -0.5 &  -0.05\\
 \hline
 $\delta$ & 0.09 & 0.1  & 0.13 \\
 \hline
\end{tabular}
\caption{List of parameters in eV for the tight-binding model given in Ref.~\cite{theuss2024single} (central columns), and used in this work (last column).}
\label{Table:parameter_set}
\end{table}

\subsection{Bogoliubov-de-Gennes Hamiltonian\label{App:HBdG}}

The Bogoliubov-de-Gennes Hamiltonian is a $16\times 16$ matrix defined as
\begin{eqnarray}
\mathcal{H}_\mathrm{BdG}(\mathbf{k})=\begin{pmatrix}
\mathcal{H}_\mathrm{TB}(\mathbf{k})\otimes\mathbb{1}_2 & \Delta(\mathbf{k})\otimes\mathbb{1}_4 \\
\Delta^\dagger(\mathbf{k})\otimes\mathbb{1}_4 & -\mathcal{H}^*_\mathrm{TB}(-\mathbf{k})\otimes\mathbb{1}_2  \\
\end{pmatrix},&
\label{Eq:BGH_model}
\end{eqnarray}
where the matrices $\mathcal{H}_\mathrm{TB}(\mathbf{k})$ and $\Delta(\mathbf{k})$ are respectively given by Eqs.~(\ref{Eq:TB_model}) and (\ref{Eq:SC_OP}).

\subsection{Method to compute the surface Green function\label{App:MethodSGF}}

For a body-centered orthorhombic lattice structure with the space group symmetry $Imm$, the reciprocal reciprocal lattice vectors are the following~\cite{bradley2009mathematical}
\begin{eqnarray}
{\bf v}_1=\dfrac{2\pi}{ca}
 \begin{pmatrix}
 c\\ 0\\a  
  \end{pmatrix},\;
{\bf v}_2=\dfrac{2\pi}{cb}
 \begin{pmatrix}
 0\\ -c\\b  
  \end{pmatrix},\;
{\bf v}_3=\dfrac{2\pi}{ba}
 \begin{pmatrix}
b\\ -a\\0 
  \end{pmatrix}~.\nonumber\\
\end{eqnarray}

To compute the surface Green function along the (0-11) plane, we first perform a basis change from the $\{k_x,k_y,k_z\}$ basis to the $\{k_x,k_\#,k_\perp\}$ basis, thanks to the rotation matrix of angle $\theta=\text{atan}(c/b)$ around the $\widehat x$-axis
\begin{eqnarray}
 \mathcal{R}_x(\theta)=
 \begin{pmatrix}
  1&0&0\\
  0&\cos\theta&-\sin\theta\\
  0&\sin\theta&\cos\theta
 \end{pmatrix}~,
\end{eqnarray}
and next we integrate over the momentum  $k_\perp$ perpendicular to the (0-11) plane. The bulk and surface spectral functions will thus be plotted as a function of~$k_x$ and~$k_\#$. The link between the elements of $({\bf k}_\parallel,{\bf k}_\perp)$ vector and $(k_x,k_\#,k_\perp)$ vector is the following: ${\bf k}_\perp=k_\perp{\bf e}_\perp$ and ${\bf k}_\parallel=k_x{\bf e}_x+k_\#{\bf e}_\#$, where $\{{\bf e}_x,{\bf e}_\#,{\bf e}_\perp\}$ is a direct orthonormal basis.

\subsection{DOS with additional d-vector contributions}\label{App:DOS_dvector}

For the sake of completeness, we calculate the DOS when adding allowed symmetry contributions to the d-vector such as
\begin{align}
  \mathbf{d}_{B_{2u}}(\mathbf{ k})=
  \begin{pmatrix}
\Delta_0\sin(ck_z)\\
C_0\sin(ak_x)\sin(bk_y)\sin(ck_z)\\
\Delta_0\sin(ak_x)
\end{pmatrix}~,
\end{align}
and
\begin{align}
  \mathbf{d}_{B_{3u}}(\mathbf{ k})=
  \begin{pmatrix}
C_0\sin(ak_x)\sin(bk_y)\sin(ck_z)\\
\Delta_0\sin(ck_z)\\
\Delta_0\sin(bk_y)
\end{pmatrix}~.
\end{align}

\begin{figure}[h!]
\begin{center}
\includegraphics[width=3.5cm]{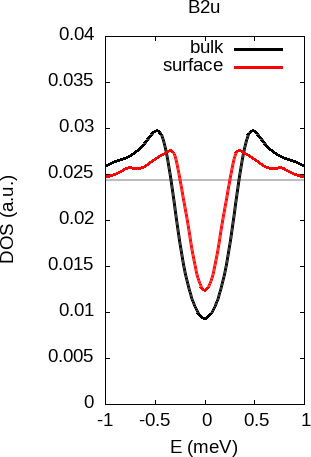}
\includegraphics[width=3.5cm]{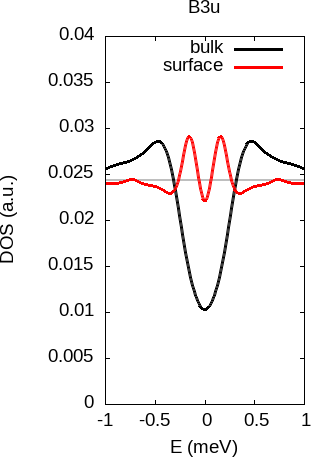}
\caption{Bulk and surface DOS for B$_{2u}$ and B$_{3u}$ pairings with $C_0=\Delta_0=0.3$ meV and $\eta=0.1$ meV.}
\label{fig:DOS_dvector}
\end{center}
\end{figure}

The result is displayed in Fig.~\ref{fig:DOS_dvector}. It shows that the peak in the B$_{3u}$ surface DOS obtained at zero energy when $C_0=0$ splits in two side-peaks when $C_0=\Delta_0$.

\section{Experimental constraints from symmetries}\label{App:Symmetry_Edge}

In this section, we present a detailed analysis of the symmetry constraints for chiral and non-chiral order parameters. As the QPI patterns depend on the orthogonality of eigenstates, certain peaks can vanish when the eigenstates are related by symmetries such as time-reversal symmetry $\mathcal{T}$. Thus implies that such considerations are important for guiding potential experimental studies. In Tables \ref{Table:Summary_Symmetry} and \ref{Table:Summary_spin}  we summarize all the relevant symmetry constraints.

\subsection{Group Theory}
Let us consider an arbitrary point group $G$ such that $P$ is generated by the following crystalline symmetries $P=\langle g_1, \cdots, g_n\rangle $ with $n$ the number of elements in the group. In the following, we only consider the space group $P_S$ which is the direct product of the translation symmetry and a point group $P$. The normal state Hamiltonian should transform as a representation $\rho$ of the space group. For $g\in P$,

\begin{align}
&\mathcal{H}_{TB}(\mathbf{k})=\rho(g)\mathcal{H}_{TB}(g\mathbf{k})\rho(g)^{-1}~.&
\end{align}

The gap function $\Delta(\mathbf{k})$ defined by Eq.~(\ref{Eq:SC_OP}) is not invariant under this transformation and instead transforms as~\cite{geier2020symmetry}
\begin{align}
&\Delta(\mathbf{k})=\rho(g)\Delta(g\mathbf{k})\rho^T(g)\Theta^*(g)~,&
\end{align}
where $\Theta(g)$ is a $1D$ representation of the symmetry group $P_S$. This means that $\Theta$ is a scalar quantity and respects the group structure. However, we can define a set of matrices $\rho_\mathcal{C}(g)$ such that the BdG Hamiltonian defined in Eq.~(\ref{Eq:BGH_model}) transforms as
\begin{align}
\label{Eq:AppEq1}&\mathcal{H}_\mathrm{BdG}(\mathbf{k})=\rho_\mathcal{C}(g) \mathcal{H}_\mathrm{BdG}(g\mathbf{k})\rho_\mathcal{C}(g)^{-1}~,&
\end{align}
and $\rho_\mathcal{C}(g)$ forms a representation of $G$. It can be checked that $\rho_\mathcal{C}$ can be expressed in terms of $\rho$ and $\Theta$ as~\cite{geier2020symmetry}
\begin{align}
&\rho_\mathcal{C}(g)=\begin{pmatrix}
\rho(g) & 0 \\
0 & \rho^*(g)\Theta(g) \\
\end{pmatrix}~.&
\end{align}

In the following, we denote $\rho$ as the representation of $G$ for the normal-state Hamiltonian, and $\rho_\mathcal{C}$ as the representation of $G$ for the BdG Hamiltonian. \\
Finally $\mathcal{H}_\mathrm{BdG}(\mathbf{k})$ possesses an additional particle-hole symmetry $\mathcal{C}$, arising from the intrinsic redundancy of the particle-hole basis~\cite{chiu2016classification}.
\begin{align}
&\mathcal{H}_\mathrm{BdG}(\mathbf{k}) = -\mathcal{C}\mathcal{H}_\mathrm{BdG}(-\mathbf{k})\mathcal{C}^{-1}~.&
\label{eq:H_C}
\end{align}

The point-group of the normal state Hamiltonian discussed in this work is $D_{2h}$~\cite{bradley2009mathematical}.
\begin{align}
&D_{2h}=\{E,m_x,m_y,m_z,C_{2x},C_{2y},\mathcal{I}\}~.&
\end{align}
The normal state is also time-reversal symmetric. Time-reversal symmetry (TRS), denoted as $\mathcal{T}$, is an anti-unitary symmetry which is spinfull here such that $\mathcal{T}^2=-1$. The symmetry group including time-reversal symmetry $\mathcal{T}$ is called a magnetic group and is given by~\cite{bradley2009mathematical}
\begin{align}
&D_{2h}^{\mathcal{T},II}=D_{2h}+\mathcal{T}D_{2h}~,&
\end{align}
where $\mathcal{T}$ acts on-site, this corresponds to a type-II magnetic point group.

The symmetry group for non-chiral order parameters is also $D_{2h}^{\mathcal{T},II}$. However, chiral order parameters exhibit lower symmetry, as they are the sum of two different representations of $D_{2h}$, with a $\pi/2$ phase difference, causing them to transform differently under crystalline symmetries and breaking on-site time-reversal symmetry $\mathcal{T}$. Their symmetry group corresponds to a type-III magnetic point group~\cite{bradley2009mathematical}. The magnetic point group for $A_u+iB_{1u}$ and $B_{2u}+iB_{3u}$ is
\begin{align}
&D_{2h}^{\mathcal{T},III}=\langle m_z, C_{2z}, \mathcal{I}, \mathcal{T}m_y, \mathcal{T}C_{2y}, \mathcal{T}m_x, \mathcal{T}C_{2}\rangle~.&
\end{align}
For $A_u+iB_{2u}$ and $B_{1u}+iB_{3u}$ it reads
\begin{align}
&D^{\mathcal{T},III}_{2h}=\langle m_y,C_{2y},\mathcal{I},\mathcal{T}m_z, \mathcal{T}C_{2z}, \mathcal{T}m_x, \mathcal{T}C_{2x}\rangle~.&
\end{align}
For $A_u+iB_{3u}$ and $B_{1u}+iB_{2u}$
\begin{align}
&D^{\mathcal{T},III}_{2h}=\langle m_x, C_{2x}, \mathcal{I}, \mathcal{T}m_y, \mathcal{T}C_{2y}, \mathcal{T}m_z, \mathcal{T}C_{2z}\rangle~.&
\end{align}

\subsection{Symmetry constraints on Green's function}

Equation~(\ref{Eq:AppEq1}) implies that the BdG Green's function defined in Eq.~(\ref{Eq:BulkGreenFunction}) transforms under a point-group transformation $g$ as
\begin{align}
&G_\mathrm{BdG}(E,\mathbf{k})=\rho(g)^{-1}G_\mathrm{BdG}(E,g\mathbf{k})\rho(g)~.&
\end{align}

Equation~(\ref{eq:H_C}) implies that under particle-hole symmetry $\mathcal{C}$, the Green's function transforms as follows
\begin{align}
&G_\mathrm{BdG}(E,\mathbf{k})=\mathcal{C}^{-1}G_\mathrm{BdG}(-E,-\mathbf{k})\mathcal{C}~.&
\end{align}

As discussed in Sec.~\ref{sec:Surface_Green}, the surface spectrum consists of two distinct contributions: one purely from the bulk which is independent of the edge impurity denoted~$G_b$, and another arising solely from edge effects named~$G_{i}$. Their expressions are given respectively in Eqs.~(\ref{Eq:BulkGreen}) and~(\ref{Eq:EdgeGreen}) such that the physical surface Green's function is the sum of these two contributions.

In the following, we consider only symmorphic crystalline symmetries, which means $\rho(g)$ is $k$-independent. The model form a representation of the space group $P_S$ which is then the semi-direct product of the $3D$ translation group $T_3$ and the point group $P=D_{2h}$, $P_S=D_{2h}\rtimes T_3$. $T_3$ is composed of the elements $t$, $t=n_1t_1+n_2t_2+n_3t_3$ where $n_i\in \mathds{N}$ and $t_i$ are elementary translations that join a point with its nearest neighbors. If we write a group element as $g^{a/b}_s=\{g^{a/b}_i|t^{a/b}\}$ with $g^{a/b}\in P$ and $t^{a/b}\in T_3$, then the semi-product $\rtimes$ means
\begin{align}
&\{g_i^a|t^a\}\{g_i^b|t^b\}=\{g_i^ag_i^b|t^a+g_i^at_b\}~.&
\end{align}

Under the unitary symmetry $g$, $G(E,\mathbf{k}_\parallel,x)$ then transforms as 
\begin{align}
\label{Eq:Appeq2}&G(E,k_x,k_\#,x)=\rho^{-1}_\mathcal{C}(g)\nonumber\\
&\times\left(\int dk_\perp e^{ixk_\perp}G_\mathrm{BdG}(E,g\mathbf{k})\right)\rho_\mathcal{C}(g)~.
\end{align}

Chiral superconductors break both time-reversal symmetry and a subset of crystalline symmetries due to their two-component nature. However, the combination of $\mathcal{T}$ and the broken crystalline symmetries $g_b$ can still form a valid symmetry, as $\mathcal{T}g_b$. This combined operator is anti-unitary, and we refer to the corresponding representation as $\rho^A$ to denote the anti-unitary nature. The corresponding transformation of $G(E,\mathbf{k}_\parallel,x)$ under these symmetries is 
\begin{align}
\nonumber&G(E,k_x,k_\#,x)=\left(\rho^A_\mathcal{C}(\mathcal{T}g_b)\right)^{-1}&\\
\label{Eq:Appeq3}&\times\left(\int dk_\perp e^{-ixk_\perp}G_\mathrm{BdG}(E,-g_b\mathbf{k})\right)\rho^A_\mathcal{C}(\mathcal{T}g_b)~.&
\end{align}

Then, for each pairing symmetry, the study of Eqs.~(\ref{Eq:Appeq2}) and~(\ref{Eq:Appeq3}) with their point group symmetry is constraining the surface spin-resolved and non-spin-resolved spectral function.

\subsection{Transformation of $G_{b}$}\label{App:bulk}

\subsubsection{Non-chiral order parameters}

Non-chiral order parameters are invariant under the $D_{2h}^{\mathcal{T},II}$ magnetic symmetry group. The unitary symmetries $m_x$ and $C_{2x}$ ensure that the bulk spectral function exhibits mirror symmetry along the $k_x$-axis and the $k_\#$-axis, respectively, for all non-chiral order parameters.

\subsubsection{Chiral order parameters}

The two chiral order parameters, $A_u+iB_{3u}$ and $B_{1u}+iB_{2u}$, are invariant under the unitary symmetries $m_x$ and $C_{2x}$. As a result, they exhibit mirror symmetry along the $k_x$-axis and the $k_\#$-axis, similar to the non-chiral order parameters. 
The other chiral order parameters break these crystalline symmetries. However, they are invariant under magnetic symmetries $\mathcal{T}C_{2x}$ and $\mathcal{T}m_x$ which  results respectively in mirror symmetry along $k_x$ and $k_\#$ axis.

\subsection{Transformation of $G_{i}$}

For the symmetry transformation of $G_{i}$, two distinct cases must be considered for each order parameter. This distinction arises from the particle-hole transformation, under which $G_\mathrm{BdG}(E,\mathbf{k})$ transforms into $G_\mathrm{BdG}(-E,-\mathbf{k})$. When $E=0$, $G_\mathrm{BdG}(E=0,\mathbf{k})\rightarrow G_\mathrm{BdG}(E=0,-\mathbf{k})$ so $G_\mathrm{BdG}(E=0,\mathbf{k})$ is constrained by particle-hole symmetry. In contrast, for $E \ne 0$, $G_\mathrm{BdG}(E,\mathbf{k})\rightarrow G_\mathrm{BdG}(-E,-\mathbf{k})$ so there is no direct constraint on $G_\mathrm{BdG}(E,\mathbf{k})$.

\subsubsection{Non-chiral order parameters}

Under $m_x$ and $C_{2x}$, $G(E,k_x,k_\#,x)$ transforms, respectively, as
\begin{align}
&G(E,k_x,k_\#,x)=\rho^{-1}_{\mathcal{C}}(m_x)G(-k_x,k_\#,x)\rho_{\mathcal{C}}(m_x)~,&\\
&G(E,k_x,k_\#,x)=\rho^{-1}_{\mathcal{C}}(C_{2x})G(k_x,-k_\#,-x)\rho_{\mathcal{C}}(C_{2x})~.&
\end{align}
However, non-chiral order parameters are also symmetric under time-reversal symmetry $\mathcal{T}$,
\begin{align}
&G(E,k_x,k_\#,x)=\rho^{-1}_{\mathcal{C}}(\mathcal{T})G(E,-k_x,-k_\#,x)\rho_{\mathcal{C}}(\mathcal{T})~.&
\end{align}
This implies that $G_{i}$ is mirror-symmetric along the $k_x$ and $k_\#$ axis, with symmetries protected by $m_x$ and $\mathcal{T}m_x$ respectively.

\subsubsection{Chiral order parameters}

We first discuss the symmetries of $G_{i}$ for chiral order parameters when $E\ne 0$. Chiral order parameters respect only a subset of the symmetries of non-chiral order parameters. We can separate the chiral order parameters that are symmetric or not under $m_x$. When they are symmetric under $m_x$, $\mathcal{T}m_x$ is broken. This is the case for $A_u+iB_{3u}$ and $B_{1u}+iB_{2u}$. In contrast, when $m_x$ is broken, $\mathcal{T}m_x$ remains a symmetry for $A_u+iB_{1u}$, $B_{2u}+iB_{3u}$, $A_u+iB_{2u}$ and $B_{1u}+iB_{3u}$. Following the discussion of non-chiral order parameters, those respecting $m_x$ are mirror symmetric along the $k_x$ axis, while order parameters respecting $\mathcal{T}m_x$ show a mirror symmetry along the $k_\#$ axis.

The case $E=0$ is left invariant under particle-hole symmetry, which implies the following relationship for every order parameters at $E=0$,
\begin{align}
&G(E=0,k_x,k_\#,x)=\mathcal{C}^{-1}G(E=0,-k_x,-k_\#,x)\mathcal{C}~.&
\end{align}
For chiral order parameters with broken $\mathcal{T}m_x$, then $\mathcal{C}m_x$ assures mirror-symmetry along the $k_\#$ axis. For chiral order parameters with broken $m_x$, then $\mathcal{C}\mathcal{T}m_x$ protects mirror-symmetry along the $k_x$ axis.

\begin{table*}[t]
\centering
\begin{tabular}{|c||c|c|c|c||c|c|c|c|c|c|}
\hline
 &$A_u$ & $B_{1u}$ & $B_{2u}$ & $B_{3u}$ & $A_u+iB_{1u}$ & $A_u+iB_{2u}$ & $A_u+iB_{3u}$ & $B_{1u}+iB_{2u}$ & $B_{1u}+iB_{3u}$ & $B_{2u}+iB_{3u}$ \\
\hline
\hline
$k_x$-bulk & $m_x$ & $m_x$ & $m_x$ & $m_x$ & $\mathcal{T}C_{2x}$ & $\mathcal{T}C_{2x}$ & $m_x$ & $m_x$ & $\mathcal{T}C_{2x}$ & $\mathcal{T}C_{2x}$ \\
\hline
$k_\#$-bulk & $C_{2x}$ & $C_{2x}$ & $C_{2x}$ & $C_{2x}$ & $\mathcal{T}m_{x}$ & $\mathcal{T}m_{x}$ & $C_{2x}$ & $C_{2x}$ & $\mathcal{T}m_{x}$ & $\mathcal{T}m_{x}$  \\
\hline
$k_x$-surface $E=0$& $m_x$ & $m_x$ & $m_x$ & $m_x$ & $\mathcal{CT}m_x$ & $\mathcal{CT}m_x$ & $m_x$ & $m_x$ & $\mathcal{CT}m_x$ & $\mathcal{CT}m_x$ \\
\hline
$k_\#$-surface $E=0$& $\mathcal{T}m_{x}$ & $\mathcal{T}m_{x}$ & $\mathcal{T}m_{x}$ & $\mathcal{T}m_{x}$ & $\mathcal{T}m_x$ & $\mathcal{T}m_x$ & $\mathcal{C}m_x$ & $\mathcal{C}m_x$ & $\mathcal{T}m_x$ & $\mathcal{T}m_x$ \\
\hline
$k_x$-surface $E\ne0$& $m_x$ & $m_x$ & $m_x$ & $m_x$ & $\times$ & $\times$ & $m_x$ & $m_x$ & $\times$ & $\times$ \\
\hline
$k_\#$-surface $E\ne0$& $\mathcal{T}m_{x}$ & $\mathcal{T}m_{x}$ & $\mathcal{T}m_{x}$ & $\mathcal{T}m_{x}$ & $\mathcal{T}m_x$ & $\mathcal{T}m_x$ & $\times$ & $\times$ & $\mathcal{T}m_x$ & $\mathcal{T}m_x$ \\
\hline
\end{tabular}
\caption{Summary of Green's function symmetries for the superconducting order parameter (not resolved in spin). 
The left column specifies the mirror symmetry plane associated with both the bulk and edge contributions to the surface Green's function. For the edge contribution the cases $E=0$ and $E\ne 0$ have to be distinguished. The top row specifies the order parameters examined in this study. Within the table, the symmetry operation responsible for protecting the mirror symmetry along the specified axis of $G_{b}$ or $G_{i}$ is listed. If a cross ($\times$) appears in place of a symmetry group element, it signifies that the mirror symmetry is not preserved.}
\label{Table:Summary_Symmetry}
\end{table*}

\begin{table*}[t]
\centering
\begin{tabular}{|c||c|c|c|c||c|c|c|c|c|c|}
\hline
 &$A_u$ & $B_{1u}$ & $B_{2u}$ & $B_{3u}$ & $A_u+iB_{1u}$ & $A_u+iB_{2u}$ & $A_u+iB_{3u}$ & $B_{1u}+iB_{2u}$ & $B_{1u}+iB_{3u}$ & $B_{2u}+iB_{3u}$ \\
\hline
\hline
$S_x$ & $\circ$ & $\times$ & $\times$ & $\circ$ & $\circ$ & $\circ$ & $\circ$ & $\circ$ & $\circ$ & $\circ$ \\
\hline
$S_y$ & $\circ$ & $\times$ & $\circ$ & $\times$ & $\circ$ & $\circ$ & $\circ$ & $\circ$ & $\circ$ & $\circ$  \\
\hline
$S_z$ & $\circ$ & $\circ$ & $\times$ & $\times$ & $\circ$ & $\circ$ & $\circ$ & $\circ$ & $\circ$ & $\circ$ \\
\hline
\end{tabular}
\caption{Summary of Green's function spin-polarization for the superconducting order parameter. A circle ($\circ$) denotes a non-vanishing spin polarization, whereas a cross ($\times$) indicates a vanishing spin polarization along a given spin direction for a given order parameter.}
\label{Table:Summary_spin}
\end{table*}

\subsection{Node structure for $B_{2u}$ and $B_{3u}$\label{App:Node_Structure}}

The node structure is shown on Fig. 21 for both B$_\text{2u}$ and B$_\text{3u}$ pairings. It can be seen in the bulk spectral function when the value of the damping $\eta$ is reduced.

\begin{figure}[h!]
\includegraphics[height=4.5cm]{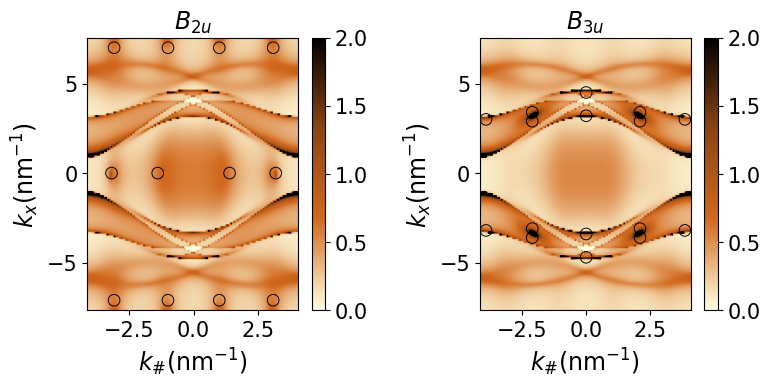}
\caption{Bulk spectral function in the (0-11)-plane  for non-chiral pairing $B_{2u}$ and $B_{3u}$ at $E=0$. The projection of the nodes onto the surface is indicated by the black circles. We take $\Delta_0=0.3$ meV and $\eta=10$ $\mu$eV.}
\label{Fig:Nodal_Bulk}
\end{figure}

\subsection{Spin-polarization\label{App:SpinPolarization}}

The normal-state Hamiltonian $h(\mathbf{k})$ used in this work is spin-degenerate and there is no spin-orbit coupling, implying that any non-trivial spin polarization in the Green's function originate from the superconducting order parameter $\Delta(\mathbf{k})$.

\subsubsection{Spin-polarization of $G_{b}$}

The electronic components of $G_{b}(E,\mathbf{k})$ contributing to  physical quantities can be written as:
\begin{eqnarray}
&&\left[G_{b}(E,\mathbf{k}_\parallel)\right]_{ee}=\int dk_\perp \left[G_\mathrm{BdG}(E,k_\perp,\mathbf{k}_\parallel)\right]_{ee}~,\\
\label{Eq:Appeq4}
&&\left[G_\mathrm{BdG}(E,\mathbf{k})\right]_{ee}=\bigg[E-h(\mathbf{k})\nonumber\\
&&-\Delta(\mathbf{k})\left[E+h^*(-\mathbf{k})\right]^{-1}\Delta^\dagger(\mathbf{k})\bigg]^{-1}~.
\end{eqnarray}
Because $h(\mathbf{k})$ is diagonal in the spin basis and $\Delta(\mathbf{k})$ and $\Delta^\dagger(\mathbf{k})$ have opposite spin polarizations, Eq.~(\ref{Eq:Appeq4}) shows that $G_{b}(E,\mathbf{k}_\parallel)$ is not spin-polarized.  Consequently, any spin polarization in $G_s$ originates exclusively from $G_{i}$.  In the following, we note $G_\Delta=\left[G_\mathrm{BdG}\right]_{eh}$ and $G_{\Delta^\dagger}=\left[G_\mathrm{BdG}\right]_{he}$. $G_\Delta$ have the polarization of $\Delta$ and $G_{\Delta^\dagger}$ of~$\Delta^\dagger$.

\subsubsection{Spin-polarization of $G_{i}$}

In the particle-hole subspace, we can write the $T$-matrix associated with the edge impurity as 
\begin{align}
&T(E,\mathbf{k}_\parallel)=\begin{pmatrix}
T_e(E,\mathbf{k}_\parallel) & T_\Delta(E,\mathbf{k}_\parallel) \\
T_{\Delta^\dagger}(E,\mathbf{k}_\parallel) & T_h(E,\mathbf{k}_\parallel) \\
\end{pmatrix}~.&
\end{align}
Similarly, one can write the Fourier transform of the single-particle Green's function as 
\begin{align}
&G(E,\mathbf{k},x)=\begin{pmatrix}
G_e^x(E,\mathbf{k}_\parallel) & G_\Delta^x(E,\mathbf{k}_\parallel) \\
G_{\Delta^\dagger}(E,\mathbf{k}_\parallel) & G_h^x(E,\mathbf{k}_\parallel) \\
\end{pmatrix}~.&
\end{align}

The electronic components of $G_{i}$ can be written as
\begin{align}
&\left[G_{i}\right]_{ee}=G^{x=1}_eT_eG^{x=-1}_e+G^{x=1}_eT_\Delta G^{x=-1}_{\Delta^\dagger}&\nonumber\\
&+G^{x=1}_\Delta T_{\Delta^\dagger} G^{x=-1}_e+G^{x=1}_\Delta T_h G^{x=-1}_{\Delta^\dagger}~.&
\end{align}

To be specific, let us consider the term $G^{x=1}_{\Delta}T_hG^{x=-1}_{\Delta^\dagger}$. Since $T_h$ is not spin-polarized, any non-trivial spin polarization must arise from $G^{x=1}_{\Delta}$ and $G^{x=-1}_{\Delta^\dagger}$. Because the polarization of $G_\Delta(E,\mathbf{k})$ and $G_{\Delta^\dagger}(E,\mathbf{k})$ is entirely determined by $\Delta(\mathbf{k})$, one can conclude that in spin-space, we have 
\begin{align}
&\vec{d}(\mathbf{k})\cdot\vec{\sigma}=\sum\limits_i d_i(\mathbf{k}) \sigma_i~,&
&G^{x}_{\Delta}\sim \sum\limits_i \alpha^x_i(k_x,k_\#) \sigma_i~,&
\end{align}
where $\alpha^x_i$ are functions of $k_x$ and $k_\#$. Importantly, $\alpha_i^{x=1}\ne \alpha_i^{x=-1}$, which enables a non-trivial spin-polarization. Finally, the polarization term can be expressed as:
\begin{align}
&G_\Delta^{x=1}T_h G^{x=-1}_{\Delta^\dagger} \sim \sum\limits_{ij} \alpha_i^{x=1}\alpha_i^{x=-1}\sigma_i\sigma_j~.&
\end{align}

The non-zero polarization terms are found to align along the $\sigma_i\sigma_j$ spin directions. Consequently, the symmetries of the spin polarization can be deduced directly from the structure of the $d$-vector, providing insight into the underlying spin-polarization patterns. 

As an example, let us study the $B_{3u}$ order parameter with $d$-vector expression given in Eq.~(\ref{eq:db3u}). In the simplest case considered in this work ${\bf d}_{B_{3u}}$ lacks a $k_x$-dependence, and the spin polarization along the $x$-direction will exhibit mirror symmetry with respect to the $x$-axis. Due to time-reversal symmetry $\mathcal{T}$, which flips the spin, the polarization will be anti-mirror symmetric along the $k_\#$-axis. This anti-mirror symmetry arises because the combined operation $\mathcal{T}m_x$ enforces the corresponding spatial transformation. More terms respecting all symmetries can be added to the $B_{3u}$ $d$-vector. The general method stays the same and the symmetry of the spin polarization can be inferred from the product of specific components of the $d$-vector. The same reasoning can also be extended to the other order parameters and explains the numerical findings.


\section{Additional plots at $E=0$}\label{App:PlotEzero}

In Sec.~\ref{Sec:Results} we presented the results for the surface spectral function, the spin-resolved surface spectral function along $\widehat x$-axis and the QPI pattern at zero-energy for various SC pairings. For the sake of completeness, in this Appendix we also provide the bulk spectral function, the spin-resolved surface spectral functions along $\widehat y$-axis and $\widehat z$-axis, and the JDOS pattern at $E=0$.

\begin{figure}[t]
\includegraphics[height=4cm]{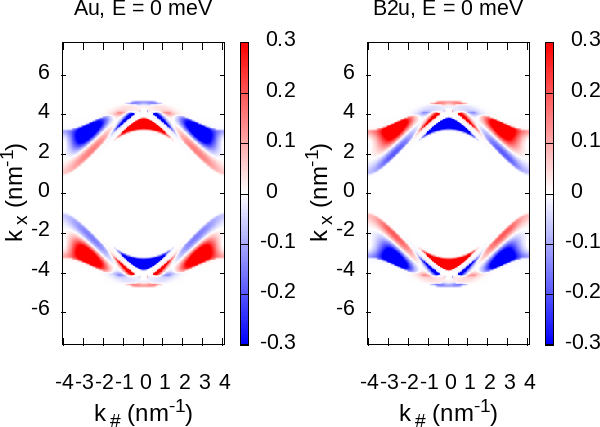}
\caption{Spin-resolved surface spectral function along $\widehat{y}$-axis for $A_{u}$ and $B_{2u}$ pairings at  $E=0$, $\Delta_0=0.3$~meV and $\eta=0.1$~meV. Note that the spin-resolved surface spectral function along $\widehat{y}$-axis is equal to zero for $B_{1u}$ and $B_{3u}$ pairings.}
\label{Fig:ASurface_E0_Sy_nonchiral}
\end{figure}

\begin{figure}[t]
\includegraphics[height=8cm]{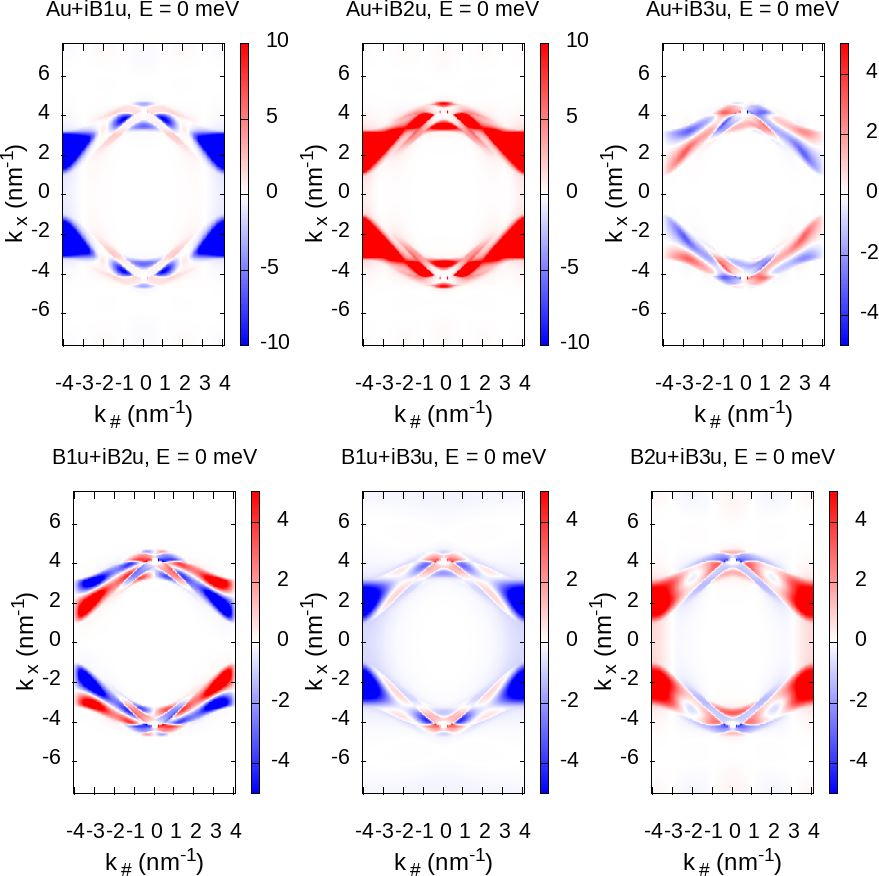}
\caption{Same as in Fig.~\ref{Fig:ASurface_E0_Sy_nonchiral} for chiral pairings.}
\label{Fig:ASurface_E0_Sy_chiral}
\end{figure}

\subsection{Spin-resolved spectral function along $\widehat{y}$- and $\widehat{z}$-axis}\label{App:SpinResolved}

\begin{figure}[t]
\includegraphics[height=4cm]{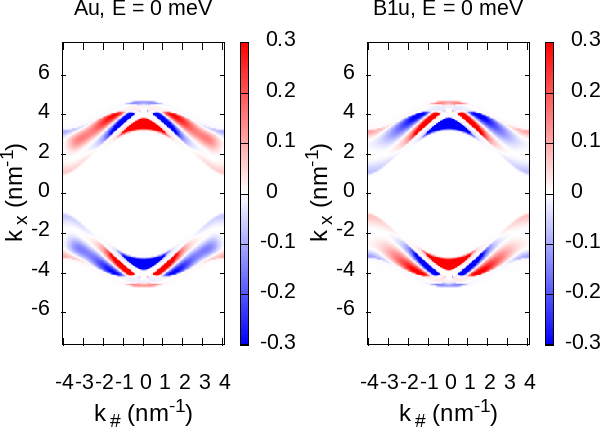}
\caption{Spin-resolved surface spectral function along $\widehat{z}$-axis for $A_{u}$ and $B_{1u}$ pairings at  $E=0$, $\Delta_0=0.3$~meV and $\eta=0.1$~meV. Note that the spin-resolved surface spectral function along $\widehat{z}$-axis is equal to zero for $B_{2u}$ and $B_{3u}$ pairings.}
\label{Fig:ASurface_E0_Sz_nonchiral}
\end{figure}

\begin{figure}[t]
\includegraphics[height=8cm]{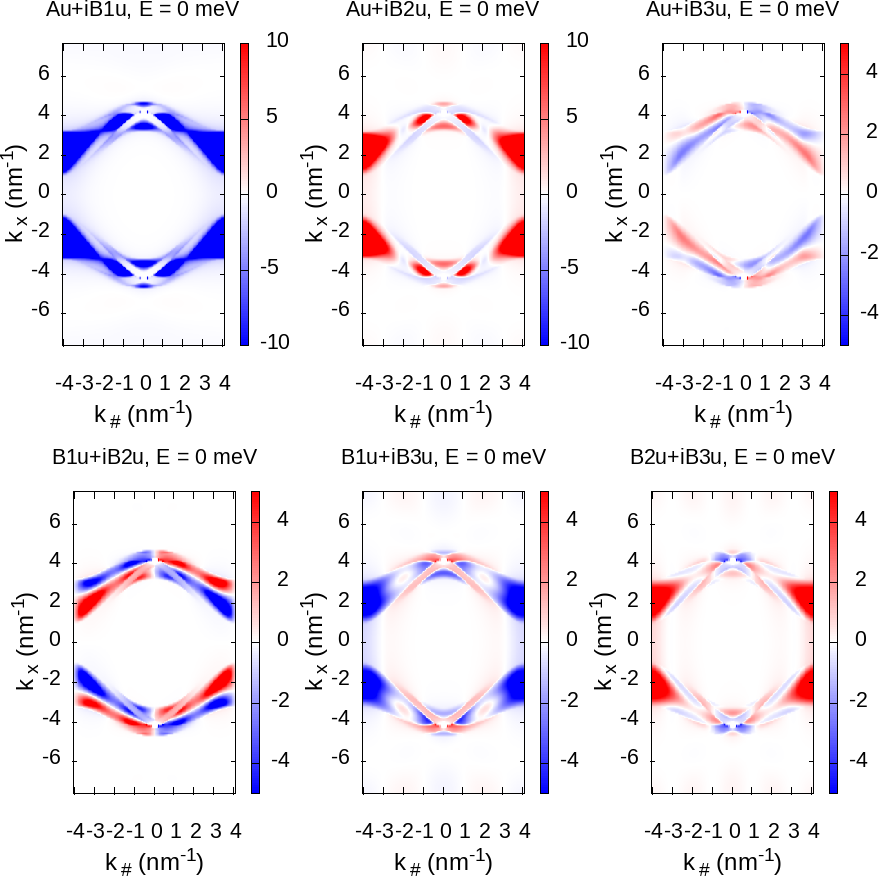}
\caption{Same as in Fig.~\ref{Fig:ASurface_E0_Sz_nonchiral} for chiral pairings.}
\label{Fig:ASurface_E0_Sz_chiral}
\end{figure}

The spin-resolved surface spectral function along $\widehat{y}$-axis and $\widehat{z}$-axis are given in Figs.~\ref{Fig:ASurface_E0_Sy_nonchiral} and \ref{Fig:ASurface_E0_Sz_nonchiral} for non-chiral pairings, and in Figs.~\ref{Fig:ASurface_E0_Sy_chiral} and \ref{Fig:ASurface_E0_Sz_chiral} for chiral pairings. We emphasize that the spin-resolved surface spectral function cancels for $B_{1u}$ and $B_{3u}$ pairings along $\widehat{y}$-axis, and for $B_{2u}$ and $B_{3u}$ pairings along $\widehat{z}$-axis. 

Non-chiral order parameters are time-reversal symmetric which imposes that surface states with opposite momenta have opposite spin polarizations, resulting in the absence of backscattering from non-magnetic impurities~\cite{hofmann2013theory}. Specifically, two states related by time-reversal symmetry satisfy $|u(\mathbf{k})\rangle =\mathcal{T}|v(-\mathbf{k})\rangle$, with $\mathcal{T}^{\,2}=-1$ in spinful systems. This implies that surface states with opposite momenta will not contribute to the QPI pattern. For chiral order parameters, time-reversal symmetry $\mathcal{T}$ is broken, so this argument does not apply directly. However, chiral order parameters still exhibit non-trivial magnetic symmetries combined with crystalline symmetries, such as $\mathcal{T}m_x$ for $A_u+iB_{1u}$ which impose constraints on the spin-resolved spectral function.

\subsection{Joint density of states\label{App:JDOS}}

\begin{figure}[t]
\includegraphics[height=8cm]{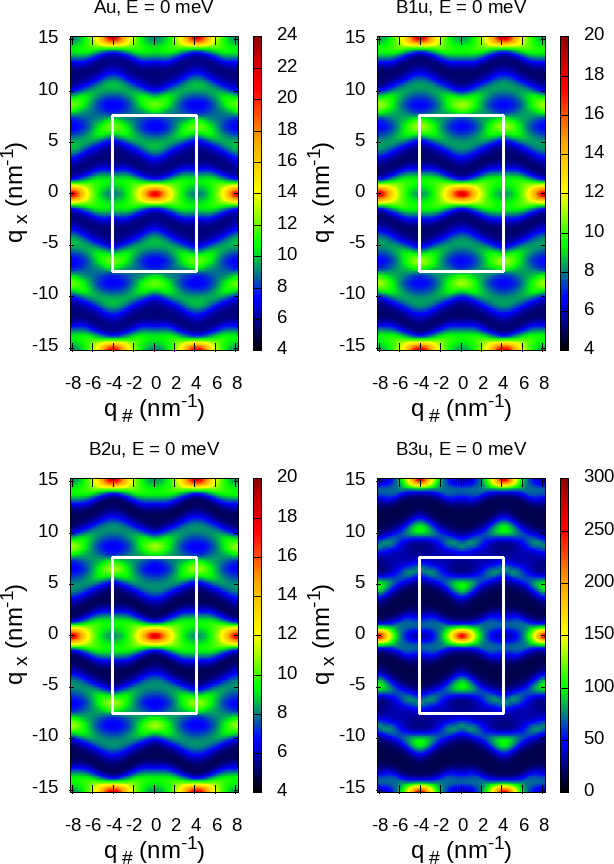}
\caption{JDOS for non-chiral pairings at $E=0$, $\Delta_0=0.3$~meV and $\eta=0.1$~meV.}
\label{Fig:Autocorrelation_E0_nonchiral}
\end{figure}

\begin{figure}[t]
\includegraphics[height=8cm]{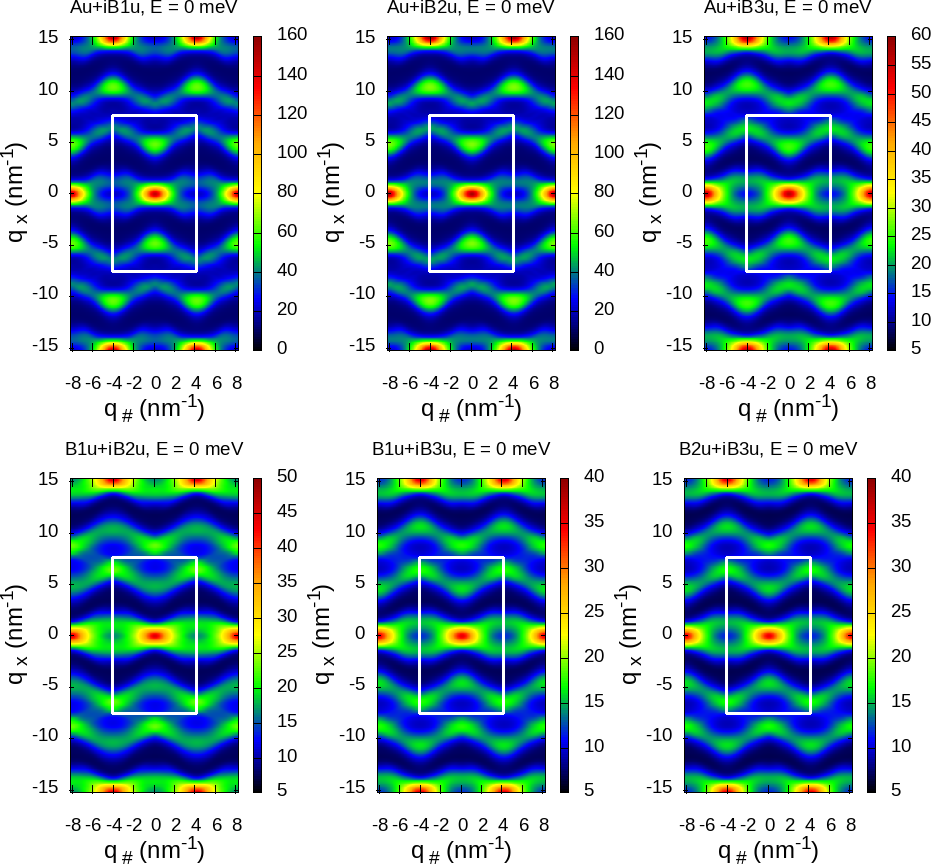}
\caption{Same as in Fig.~\ref{Fig:Autocorrelation_E0_nonchiral} for chiral pairings.}
\label{Fig:Autocorrelation_E0_chiral}
\end{figure}

The JDOS provides an intuitive picture of possible peak locations in the QPI pattern. In Figs.~\ref{Fig:Autocorrelation_E0_nonchiral} and ~\ref{Fig:Autocorrelation_E0_chiral} we shown the JDOS at $E=0$. The observed peaks correspond to wavevectors connecting two regions of high intensity in the surface spectral function. From the bulk and edge contribution to the surface spectral function, shown respectively in Figs.~\ref{Fig:A0_nonchiral}, \ref{Fig:A0_chiral}  and Figs.~\ref{Fig:ASurface_E0_nonchiral}, \ref{Fig:ASurface_E0_chiral}, we observe that the spectral function tends to show similar regions of high intensity for all the OPs. This implies that the peaks in the autocorrelation spectrum  will share roughly the same wavevectors.
This shows that the autocorrelation spectrum does not provide sufficient information to differentiate between different OPs for this system. However, the experimental measurements should not be compared with the JDOS but with the results of the full T-matrix calculations (the QPI patterns). In the full calculations due to phase cancellations the intensity and the position of the peaks can be very different from one OP to the next, thus the comparison with the experiments allow us to identify the correct OP.

\section{Study of energy dependence\label{App:E_Plot}}

In this Appendix we study how the surface spectral function and the QPI patterns change with increasing energy~$E$ from 0 meV to 0.25 meV, i.e, within the superconducting gap equal to $\Delta_0=0.3$ meV.

\subsection{Surface energy band\label{App:band_ABCD}}

\begin{figure}[t]
\includegraphics[width=3cm]{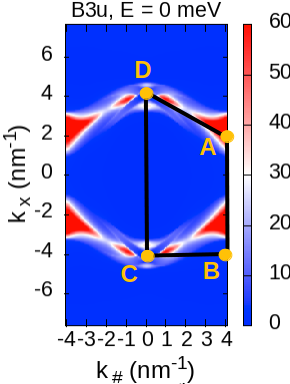}
\caption{Path ABCDA in the (0-11)-plane.}
\label{Fig:band_ABCD}
\end{figure}

\begin{figure}[t]
\includegraphics[width=7cm]{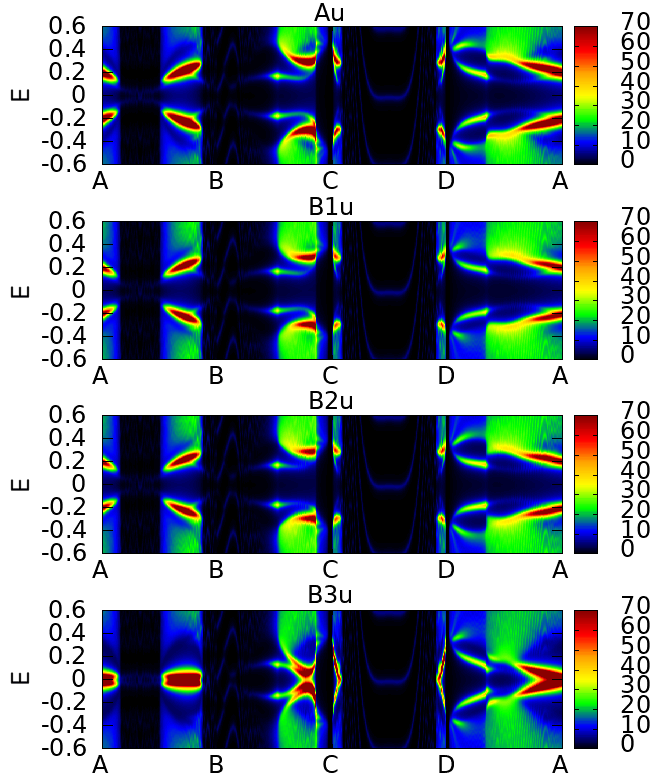}
\caption{Energy band dispersion (in meV) along the path ABCDA shown in Fig.~\ref{Fig:band_ABCD} for non-chiral pairings at  $\Delta_0=0.3$~meV and $\eta=0.03$~meV.}
\label{Fig:band_ABCD_nonchiral}
\end{figure}

\begin{figure}[t]
\includegraphics[width=7cm]{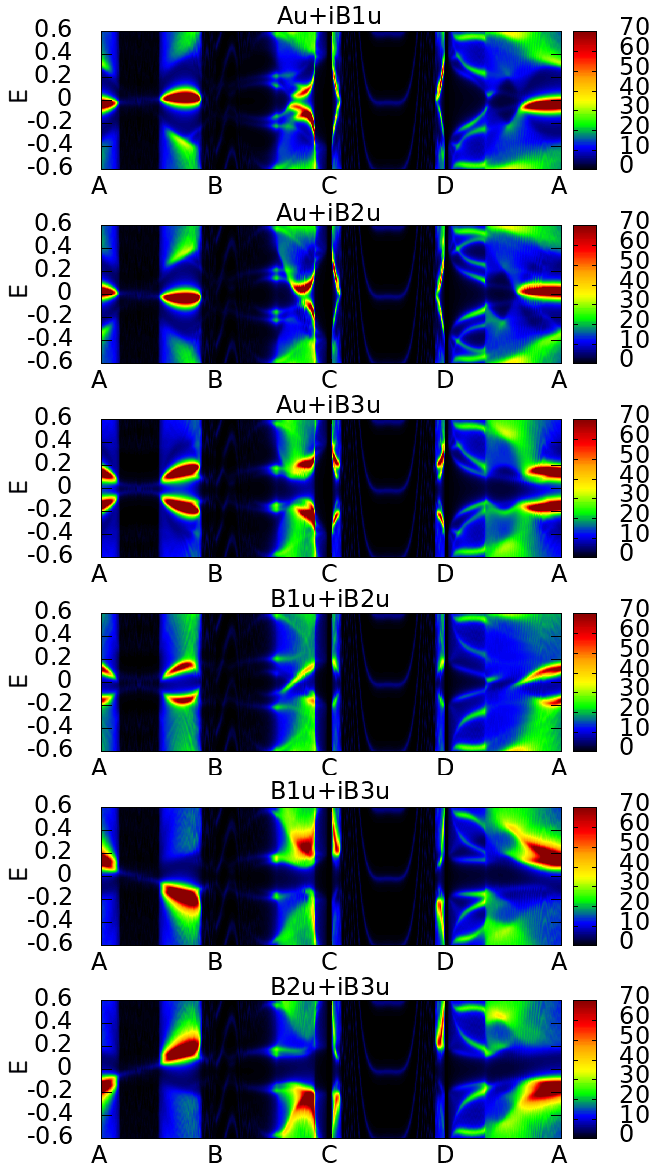}
\caption{Same as in Fig.~\ref{Fig:band_ABCD_nonchiral} for chiral pairings.}
\label{Fig:band_ABCD_chiral}
\end{figure}

Figures~\ref{Fig:band_ABCD_nonchiral} and \ref{Fig:band_ABCD_chiral} show the surface energy band along a path ABCDA on the (0-11)-plane depicted in  Fig.~\ref{Fig:band_ABCD}, at $\Delta_0=0.3$~meV and $\eta=0.03$~meV. This path is chosen to highlight the main characteristics, in particular the high intensity features in the surface energy band. We took a smaller value for the damping ($\eta=\Delta_0/10$) compared with the one used for the surface spectral function and QPI ($\eta=\Delta_0/3$) to better see the dispersing bands.

\subsection{Surface spectral function at $E\ne 0$\label{App:A_E}}

\begin{figure}[t]
\includegraphics[width=14cm]{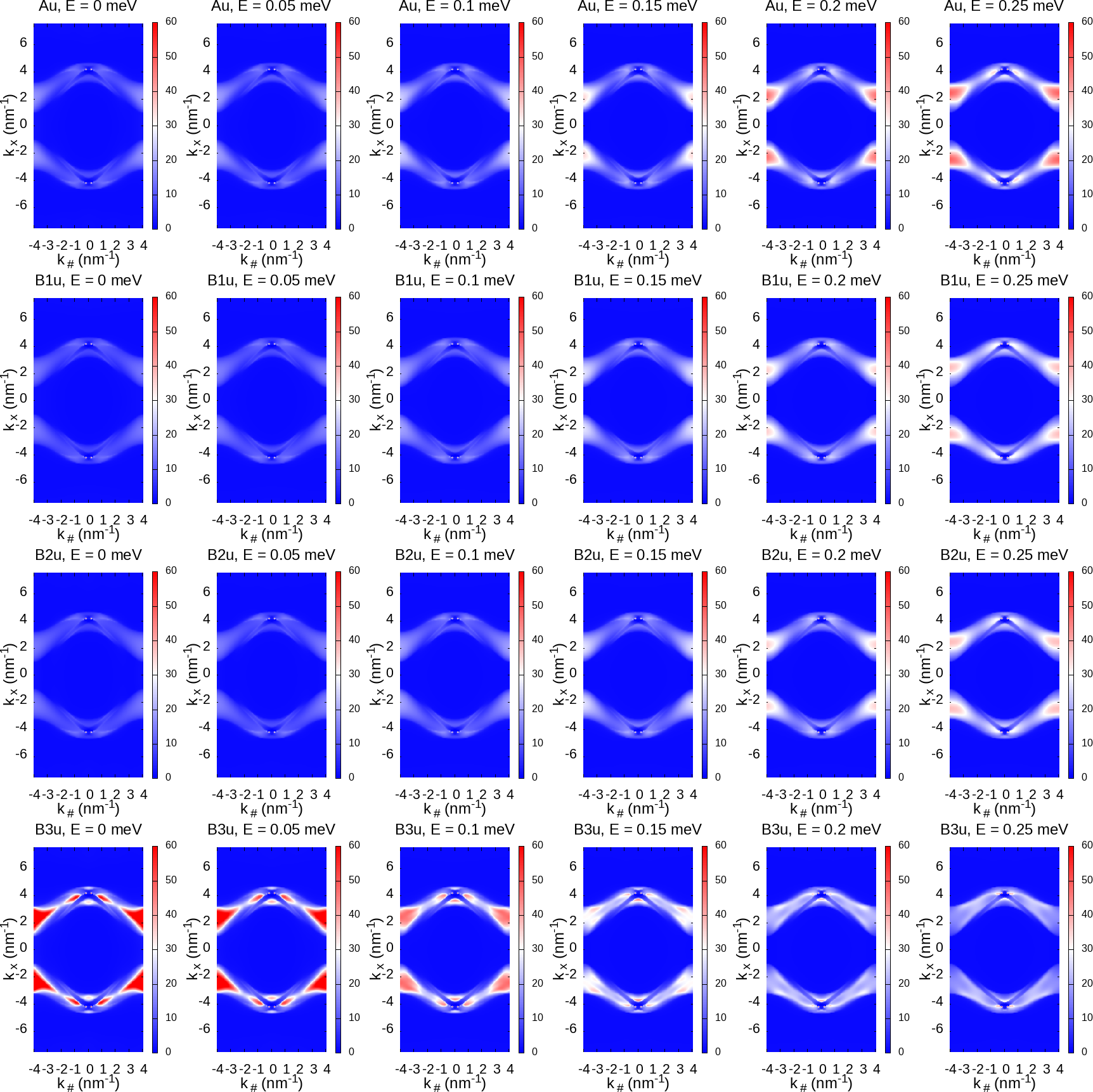}
\caption{Surface spectral function for non-chiral pairings at different energies. We take $\Delta_0=0.3$ meV and $\eta=0.1$ meV.}
\label{Fig:ASurface_E_nonchiral}
\end{figure}

\begin{figure}[t]
\includegraphics[width=14cm]{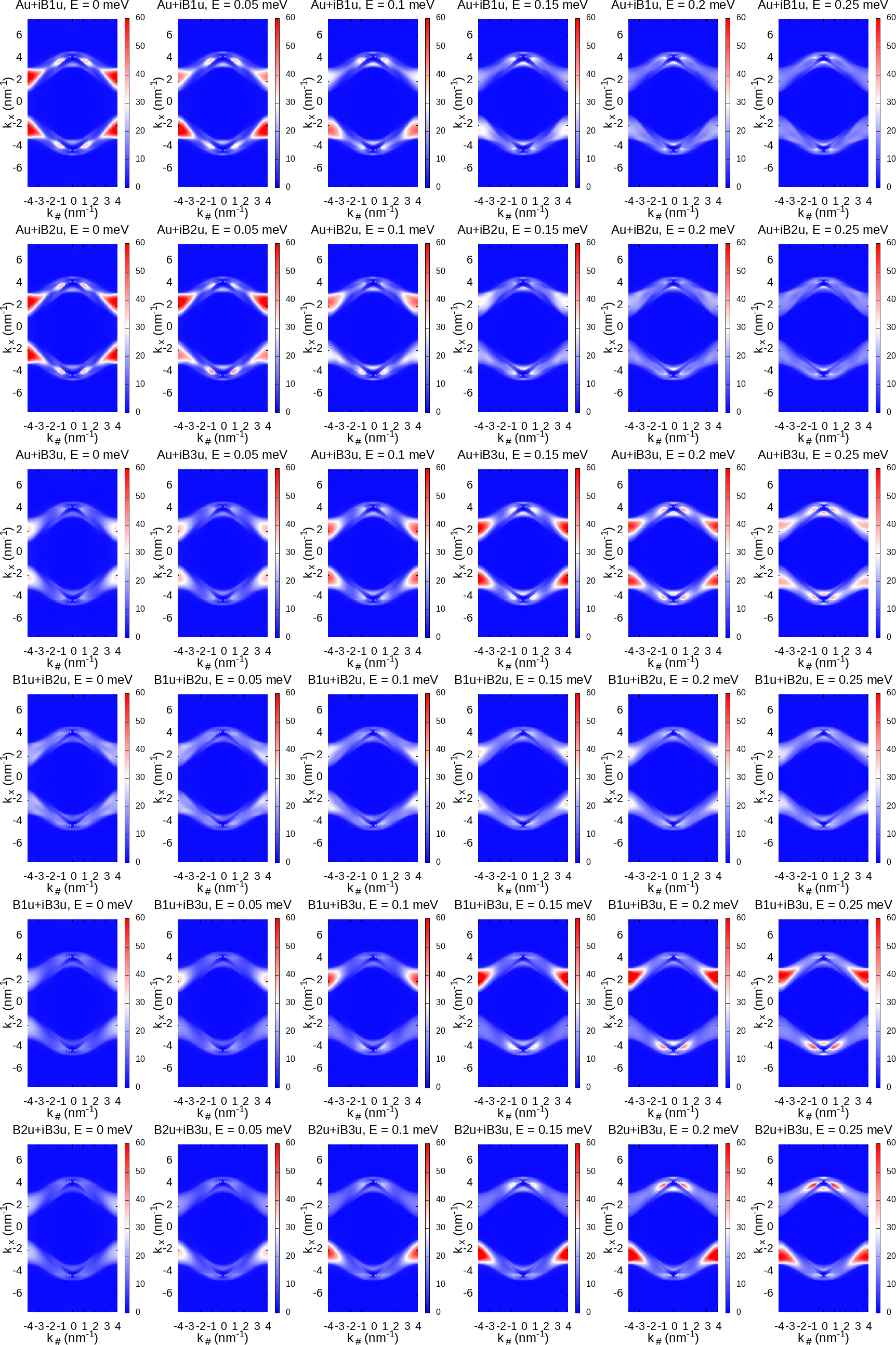}
\caption{Same as in Fig.~\ref{Fig:ASurface_E_nonchiral} for chiral pairings.}
\label{Fig:ASurface_E_chiral}
\end{figure}

Figures~\ref{Fig:ASurface_E_nonchiral} and \ref{Fig:ASurface_E_chiral} depict the evolution of the surface spectral function for an energy varying from 0 to 0.25 meV. We note that, as also discussed in the main text, the amplitude of the surface spectral function is highly dependent on energy, thus superconducting surface states visible at low energy can disappear at higher energy for some pairings, or inversely, superconducting surface states absent at low energy can appear at higher energy for others. 

We also note that at $E\ne 0$ the mirror-symmetry with respect to the $k_x$-axis and $k_\#$-axis is preserved for non-chiral pairings, but can be broken for some of the chiral pairings. The explanation for this is that, although the triplet order parameters associated with $A_u$, $B_{1u}$, $B_{2u}$ and $B_{3u}$ pairings are not invariant under the spatial symmetries of the group $Immm$, they remain invariant under symmetries in the superconducting phase, because a gauge transformation can remove a $U(1)$ phase~\cite{ono2020refined}. However, when two order parameters that transform differently under a crystalline symmetry~$g$ are combined, such as in $B_{1u}+iB_{2u}$ for example, gauge invariance alone is insufficient to restore the crystalline symmetry~$g$. The symmetry breaking for the chiral order parameters is reflected in the surface spectral function plots: as shown in Figs.~\ref{Fig:ASurface_E_nonchiral} and \ref{Fig:ASurface_E_chiral}, the edge contribution to the surface spectral function for $E\ne 0$ remains mirror-symmetric for the non-chiral pairings, while this symmetry is broken in the chiral case.

\subsection{QPI patterns at $E\ne 0$\label{App:E_QPI}}

\begin{figure}[t]
\includegraphics[width=14cm]{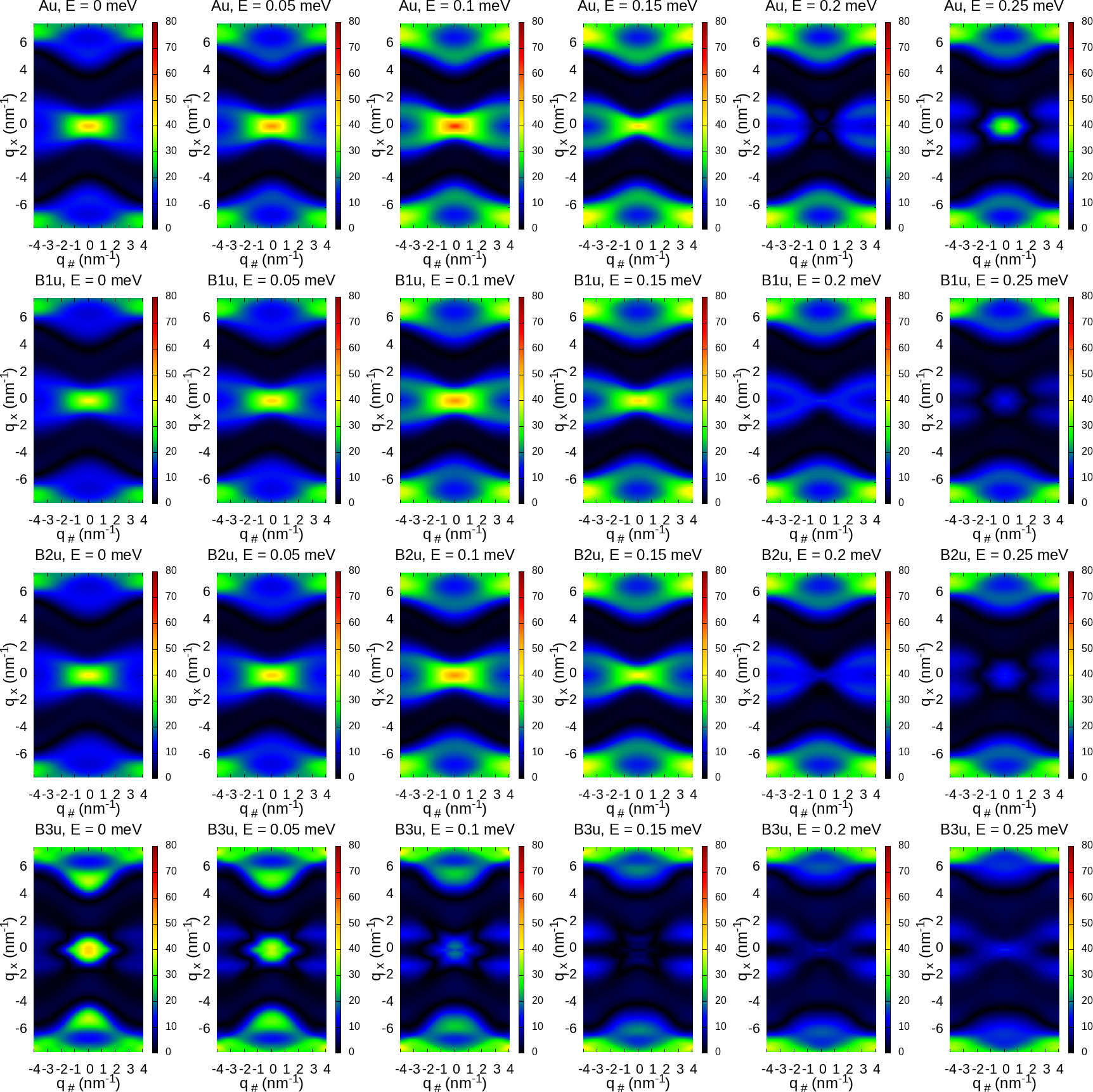}
\caption{QPI pattern for non-chiral pairings at different energies. For better visibility here we restrict the plots to the first Brillouin zone. We take $\Delta_0=0.3$ meV and $\eta=0.1$ meV.}
\label{Fig:QPInonchiral_ne0}
\end{figure}

\begin{figure}[t]
\includegraphics[width=14cm]{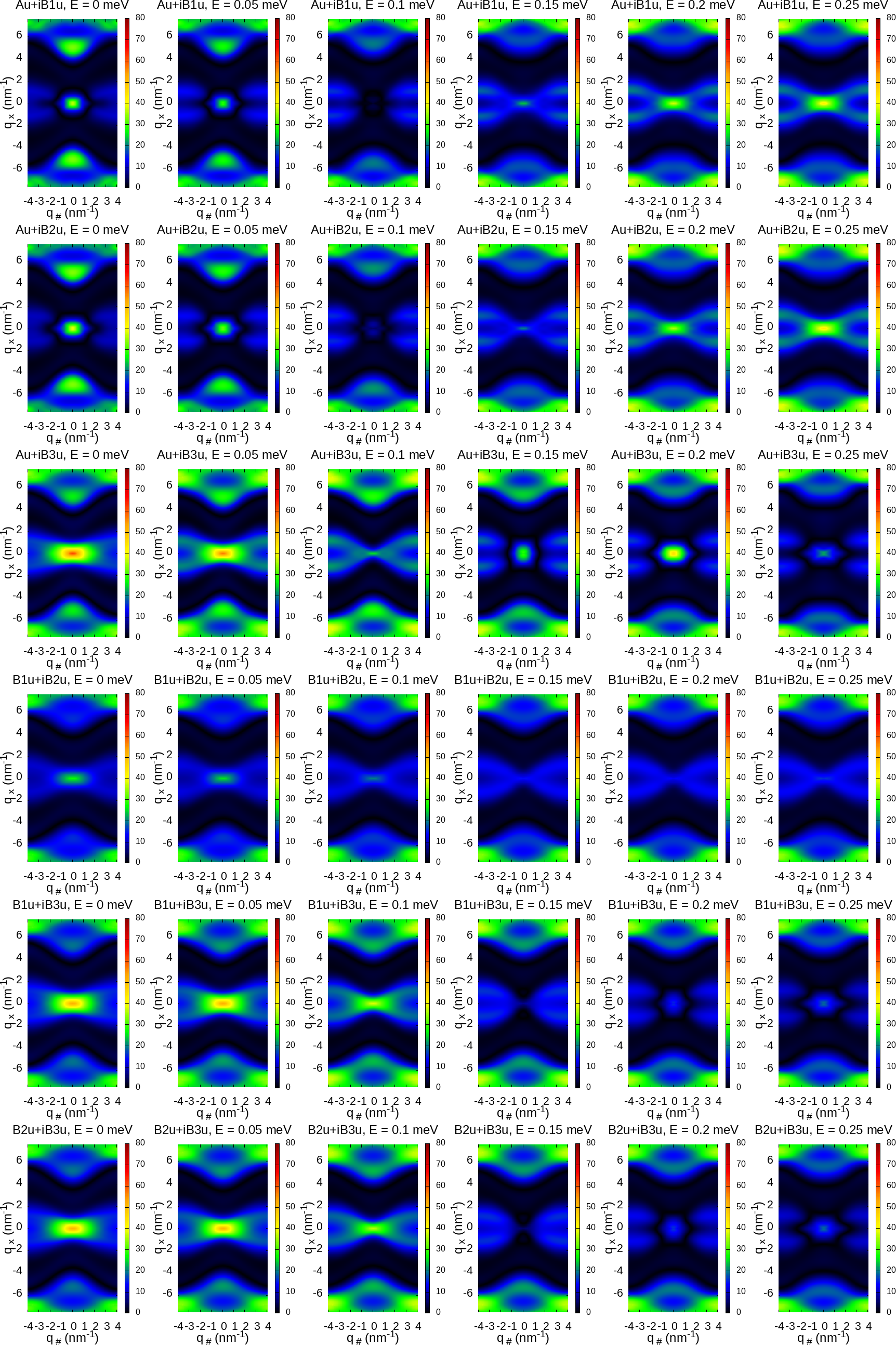}
\caption{Same as in Fig.~\ref{Fig:QPInonchiral_ne0} for chiral pairings.}
\label{Fig:QPIchiral_ne0}
\end{figure}

Figures~\ref{Fig:QPInonchiral_ne0} and \ref{Fig:QPIchiral_ne0} describe the evolution of the QPI pattern for an energy varying from 0 to 0.25 meV. Note that the ${\bf q}_1$ peak visible for $B_{3u}$, $A_u+iB_{1u}$, $A_u+iB_{2u}$, $A_u+iB_{3u}$, $B_{1u}+iB_{3u}$ and $B_{2u}+iB_{3u}$  pairings is stable in position and remains visible up to $E\approx 0.15$ meV.


\bibliography{UTe2.bib}


\pagebreak

\onecolumngrid

\twocolumngrid

\end{document}